\documentclass[12pt,aps,nofootinbib,preprintnumbers,groupedaddress,letterpaper]{article}

\usepackage{hyperref}
\usepackage[dvipsnames]{xcolor}
\usepackage[normalem]{ulem}
\usepackage{amsmath, bbold}
\usepackage{enumerate}
\usepackage{amsfonts}
\usepackage{yfonts}

\usepackage{cite}

\usepackage{graphicx}
\usepackage{graphbox}
\usepackage{subcaption}

\newcommand\comment[1]{}
\newcommand\moebius{M\" obius }
\newcommand\moebiuss{M\" obius}

\newcommand\schr{Schr\" odinger }

\newcommand\poincare{Poincar\' e }

\newcommand\ov{\over }

\def\le{\left}
\def\ri{\right}

\def\({\left(}
\def\){\right)}
\def\<{\langle}
\def\>{\rangle}

\def\tr{\mathop{\rm tr}}

\newcommand\half{{\ensuremath{\frac{1}{2}}}}
\newcommand\p{\ensuremath{\partial}}

\newcommand\field[1]{{\ensuremath{\mathbb{{#1}}}}}

\newcommand{\RR}{\field{R}}

\newcommand{\ZZ}{\field{Z}}

\newcommand{\be}{\begin{equation}}
\newcommand{\ee}{\end{equation}}
\newcommand{\bea}{\begin{eqnarray}}
\newcommand{\eea}{\end{eqnarray}}
\newcommand{\bwt}{\begin{widetext}}
\newcommand{\ewt}{\end{widetext}}

\newcommand{\bi}{\begin{itemize}}
\newcommand{\ei}{\end{itemize}}
\newcommand{\ben}{\begin{enumerate}}
\newcommand{\een}{\end{enumerate}}
\newcommand{\bca}{\begin{cases}}
\newcommand{\eca}{\end{cases}}
\newcommand{\bln}{\begin{align}}
\newcommand{\eln}{\end{align}}
\newcommand{\bst}{\begin{split}}
\newcommand{\est}{\end{split}}

\renewcommand{\Im}{\textrm{Im}\,}

\newcommand\Deltax{M}

\def\Xint#1{\mathchoice
{\XXint\displaystyle\textstyle{#1}}%
{\XXint\textstyle\scriptstyle{#1}}%
{\XXint\scriptstyle\scriptscriptstyle{#1}}%
{\XXint\scriptscriptstyle\scriptscriptstyle{#1}}%
\!\int}
\def\XXint#1#2#3{{\setbox0=\hbox{$#1{#2#3}{\int}$}
\vcenter{\hbox{$#2#3$}}\kern-.5\wd0}}

\def\dashint{\Xint-}

\begin{document}

\begin{titlepage}

\begin{flushright}
QMUL-PH-23-01\\
\end{flushright}

\vspace{5mm}

  \begin{center}

\centerline{\Large \bf {The 't Hooft equation as a quantum spectral curve}}

\bigskip
\bigskip

{\bf David Vegh}

\bigskip

\small{
{ \it   Centre for Theoretical Physics, Department of Physics and Astronomy \\
Queen Mary University of London, 327 Mile End Road, London E1 4NS, UK}}

\medskip

{\it email:} \texttt{d.vegh@qmul.ac.uk}

\medskip

{\it \today}

\bigskip


\begin{abstract}

In an attempt to establish a link between different quantization methods, I examine the massless~'t~Hooft equation, which governs meson bound state wavefunctions in 2d  $SU(N)$ gauge theory in the large-$N$ limit. The integral equation can also be obtained by lightcone quantizing a folded string in flat space. The folded string is a  limiting case of a more general setup: a four-segmented string moving in~AdS$_3$.  I  compute its classical spectral curve  by using celestial variables and planar bipartite graphs (on-shell diagrams/brane tilings). The adjugate of the Kasteleyn matrix vanishes at two special points, which ensures that the string segments form a closed loop in target space. The Hamiltonian takes on a Ruijsenaars-Schneider form and the phase space is a coadjoint $SL(2)$ orbit whose middle region has been removed. In AdS, the~'t~Hooft equation acquires an extra term, which has  previously been proposed as an effective confining potential in QCD. After an integral transform, the  equation can be inverted in terms of a finite difference equation. I show that this difference equation can be interpreted as the quantized (non-analytic) spectral curve of the string. I calculate the spectrum numerically, which interpolates between  $\Deltax^2=n(n+1)$ in the tensionless limit and 't~Hooft's nearly linear Regge trajectory at infinite AdS radius.

\end{abstract}

\end{center}

\end{titlepage}

\vskip-1.5cm
\tableofcontents

\clearpage

\section{Introduction}

Quantizing a classical system is not a straightforward procedure and the prescription can fail for several reasons. In general, it is difficult to find a Hilbert space and assign self-adjoint operators to the relevant observables. Operators typically do not commute, consequently an ordering ambiguity arises and various quantization schemes may be used to find a consistent quantum theory. Especially difficult is the quantization of constrained systems and systems with gauge redundancies.
An example for such a system is the bosonic (Nambu-Goto) string  on a general background. On highly symmetrical target spaces the worldsheet theory may be integrable---at least classically \cite{Pohlmeyer:1975nb}. Two-dimensional integrable theories can oftentimes be discretized in an exact manner (so that both spatial and time directions are discrete), which renders the phase space finite dimensional. This motivates the study of discrete strings, in the hope that perhaps the quantization is under more control\footnote{Another motivation might come from the fact that some sort of discretization scheme is necessary when the string motion is simulated on a computer. String-like excitations can also appear in a statistical physics model that lives on a discrete lattice, which further motivates the study of discrete strings. See \cite{Giles1977, Klebanov:1988ba} for earlier works on discrete strings.}.

In the theory of integrability, Baxter's T--Q equation \cite{Baxter:1972hz, Baxter:1972wg} plays a key role in various models. The equation is a finite difference equation\footnote{Finite difference equations also play an important role in other areas of physics, such as in the diffraction theory of waves incident on wedges or cones  (see e.g. \cite{Antipov}), or in the quantization of a self-gravitating spherically symmetric dust shell in comoving coordinates \cite{Hajicek92, berezin97}.} and in certain cases it can be interpreted as a quantized spectral curve.
Recently, integrability techniques have been brought to bear on the AdS/CFT correspondence \cite{Maldacena:1997re, Gubser:1998bc, Witten:1998qj}  in the planar limit. In the seminal paper \cite{Minahan:2002ve} it was recognized that the one-loop spectrum of $\mathcal{N}=4$ SYM is captured by an integrable spin chain.
It was also discovered that superstrings on the dual AdS$_5\times S^5$  background are classically integrable \cite{Bena:2003wd}.
These results generated a great deal of interest, which culminated in the construction of the AdS$_5$/CFT$_4$ quantum spectral curve\footnote{Recently there have been proposals for the quantum spectral curve of AdS$_3\times S^3 \times T^4$ \cite{Cavaglia:2021eqr, Ekhammar:2021pys}.} \cite{Gromov:2013pga, Gromov:2014caa} (see \cite{Gromov:2017blm, Kazakov:2018ugh, Levkovich-Maslyuk:2019awk} for
reviews of the subject).

\begin{figure}[h]
\begin{center}
\includegraphics[width=8cm]{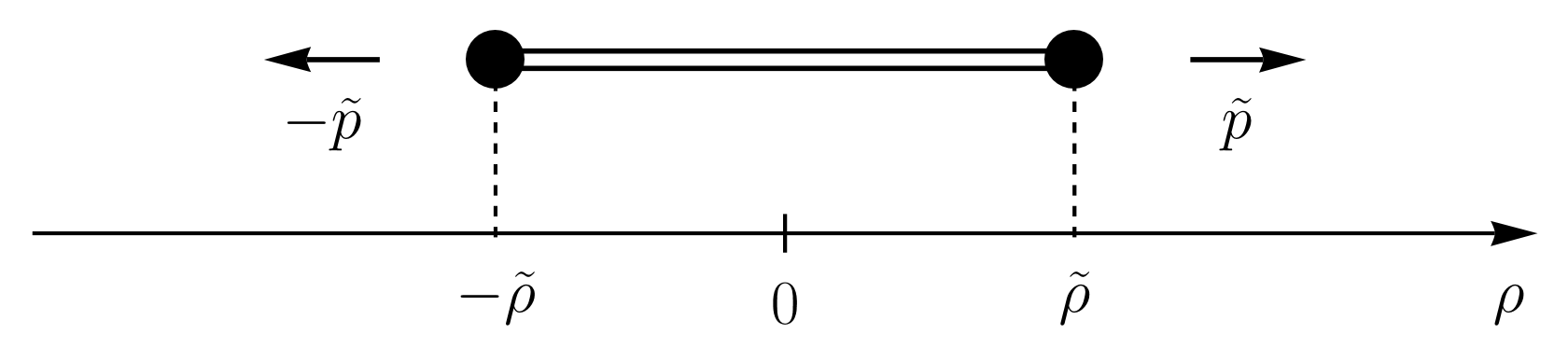}
\caption{\label{fig:yoyo}
A folded ``yoyoing" string consists of two massless particles connected by two string segments. The particles move with the speed of light in AdS and bounce back when they reach maximum extension. The motion is periodic in time. The position and momentum of one of the particles is denoted by $\tilde \rho$ and $\tilde p$, respectively.
}
\end{center}
\end{figure}

The existing quantum spectral curve constructions are rather intricate  due to the large symmetry groups. In this paper we will explore a much simpler system: a closed segmented string and make a first step towards comparing different approaches to quantization. The string under consideration is the folded string, which is depicted in Figure~\ref{fig:yoyo}. It is composed of two  massless particles connected by two identical string segments so that the string forms a (collapsed) closed loop in flat space. In a (1+1)-dimensional target space the system has a four-dimensional phase space, but the center-of-mass coordinates will decouple leaving us with a two-dimensional phase space.
Note that fixing the number of particles in a relativistic quantum theory is dangerous as it generally leads to a broken Lorentz symmetry. Nevertheless, by going to an infinite momentum frame one can deal with this issue and obtain a consistent quantum theory.

In order to compare the variables used in lightcone quantization with those used to parametrize the spectral curve, we first need to compute the spectral curve itself.
Since the calculation in flat space is somewhat degenerate, we compute the curve in AdS$_3$ instead.
We find a transformation between physical coordinates and the quasimomentum and the spectral parameter, denoted by $p$ and $u$, respectively. The physical configuration space turns out to be the $|u| \ge 2g$ subset of the real line, where $g:={L^2 /(2 \pi \alpha')}$ is a constant and $L$ is the AdS radius. By cutting out the unphysical $(-2g, +2g)$ region, and denoting the new coordinate by $q$, we obtain a non-analytic real spectral curve,
\be
   \label{eq:egyes}
     e^p + e^{-p}  + 2 - {\Deltax^2 \ov {  q^2 }+ 4g |q|}  = 0 \, ,
\ee
where $\Deltax$ is the energy of the string.
We recall the derivation of the 't Hooft equation by lightcone quantizing the folded string and generalize it to the case of AdS. The only contribution coming from the curvature of the target space is an extra term in the equation.  We find a canonical transformation between $(p,u)$ and the ligthcone variables, which (following \cite{Fateev:2009jf}) enables us to invert the modified 't Hooft equation and obtain the difference equation,
\be
  \nonumber
  Q(q+i)+Q(q-i) -2 Q(q) = -  {\Deltax^2   \ov {  q^2  }+ 4g q   \coth({\pi q) }  } Q(q)  \, ,
\ee
which is identified as the quantum version of the non-analytic spectral curve in \eqref{eq:egyes}.

The  paper is organized as follows.  In section 2 we discuss the folded string in  AdS$_3$. This is a one-dimensional example and its Hamiltonian can be computed in a straightforward way. The section closely follows section 4.3 in \cite{Callebaut:2015fsa}. In section 3 we discuss more general segmented strings, in particular the four-segmented case, which will be at the center of our focus. The embedding into AdS$_3$ is described and the target space string energy is calculated. Section 4 discusses the spectral curve of the four-segmented string. The calculation of the curve is done in different ways:  by using  $2 \times 2$ Lax matrices containing (i) celestial and  (ii) \moebius invariant tiling variables. In order to ensure that the string closes in the target space, certain ``closing constraints'' must be satisfied, which are discussed in some detail. Section~5 introduces canonical variables on the spectral curve and  derives the $SL(2)$ Ruijsenaars-Schneider Hamiltonian for the four-segmented string. We discuss the transformation between various coordinates and compute their Poisson brackets. In section 6 we recall the derivation of the 't Hooft equation by lightcone quantizing the string and find the map between lightcone variables and spectral curve coordinates. We discuss the integral transform of the original 't Hooft equation along with its AdS counterpart. Section 7 provides  details on the numerical calculation of the spectrum. In the appendix we point out a potentially interesting reformulation of the closing constraints in terms of the adjugate of the Kasteleyn matrix.

 \clearpage

\section{The folded string  in AdS$_2$}

\label{sec:two}

Strings with longitudinal kink modes have been studied since the early days of string theory.
These can be thought of as a set of  particles connected by strings.
For a non-comprehensive list of papers  discussing strings with longitudinal degrees of freedom in flat space, we refer the reader to  \cite{Patrascioiu:1974un, PhysRevD.13.2364, Artru:1979ye, Soederberg}.
The idea of a folded string with several folds can be generalized to a $d$-dimensional anti-de Sitter (AdS$_d$) target space. In this section we will discuss the simplest folded string, which moves in AdS$_2$ and has two folds.  The system can be depicted as two particles connected by two identical string segments as shown in Figure \ref{fig:yoyo}. In this section, we summarize the results of section 4.3 in \cite{Callebaut:2015fsa}, where the authors  derived the spacetime Hamiltonian using a direct approach.

The action can be written as \cite{Ficnar:2013wba}
\be
  \nonumber
  \mathcal{S}_\textrm{string} = -{1 \over 4\pi\alpha'} \int_M d^2 \sigma \, \sqrt{-h} h^{ab}
    \partial_a X^\mu \partial_b X^\nu G_{\mu\nu} +
   \int_{\partial M} d\xi \, {1 \over 2\eta} \dot{X}^\mu \dot{X}^\nu G_{\mu\nu} \,,
\ee
where $M$ is the worldsheet and  $\partial M$ is its boundary. The endpoints carry momentum, which is taken into account by the second term. The worldsheet metric $h_{ab}$ is  determined  by its equation of motion (up to a conformal factor).  $\eta=\eta(\xi)$ is the einbein on the boundary worldline, parametrized by the coordinate $\xi$.
The AdS$_2$ target space metric is given by
\be
  \nonumber
    ds^2 = L^2 (-\cosh^2 \rho \, d\tau^2 + d\rho^2 ) \, ,
\ee
where the domain of both coordinates is the entire real line.
Following \cite{Callebaut:2015fsa}, we parametrize the
worldsheet using the $\tau$ and $\rho$ coordinates. For simplicity, we take a symmetric string configuration centered at $\rho=0$ and
parametrize the worldline of the endpoints using the coordinate $\xi=\tau$.
The location of one of the endpoints is denoted by $\tilde \rho(\tau)$  and it will be the dynamical variable in the following. Thus, $\rho$ runs from $-|\tilde \rho|$ to $+|\tilde \rho|$, covering half of the worldsheet, i.e. one of the string segments in Figure \ref{fig:yoyo}.
The string Lagrangian can be computed explicitly \cite{Callebaut:2015fsa}
\be
   \mathcal{L} =  - {2 L^2\ov \pi \alpha'} \int_{0}^{| \tilde \rho | } d\rho \, \cosh \rho + {L^2 \over \eta }(-\cosh^2 \tilde \rho + \dot{ \tilde \rho}^2 )  = - {2  L^2\ov \pi \alpha'} \sinh | \tilde \rho | + {L^2 \over \eta }(-\cosh^2 \tilde \rho + \dot{ \tilde \rho}^2 ) \, ,
\ee
Performing the Legendre transformation and eliminating $\eta$ gives the Hamiltonian in terms of the location of one of the endpoints $\tilde\rho(\tau)\in (-\infty, \infty)$ and the conjugate momentum $\tilde p(\tau)$  \cite{Callebaut:2015fsa}
\be
  \label{eq:hami}
 \tilde H(\tilde p, \tilde \rho) = \tilde p \dot{\tilde \rho} - \mathcal{L} =  |\tilde{p}| \cosh \tilde \rho + {2 L^2\over \pi\alpha'} \sinh |\tilde \rho |\, ,
\ee
and we have $\{\tilde \rho, \tilde{p} \} =  1 $.
Note that in the $\tilde \rho \ll 1$ flat space  limit, \eqref{eq:hami} reduces to the ``zigzag'' Hamiltonian \cite{Donahue:2019adv, Donahue:2019fgn, Donahue:2022jxu}, which for two particles is given by
\be
  \label{eq:zigzag}
 \tilde H_\textrm{flat} =    |\tilde{p}|  + {2 L^2\over \pi\alpha'}   |\tilde \rho |=    |\tilde{p}|  + 4g  |\tilde \rho |\, ,
\ee
where we have introduced the constant
\be
  \nonumber
  g := {L^2 \ov   2 \pi \alpha'} \, .
\ee
The authors of  \cite{Callebaut:2015fsa} also write down the WKB quantization condition,
\be
  \nonumber
  {\mathcal{S} \over 4} \equiv   \int_0^{\tilde\rho_0} d\tilde\rho \, \tilde p(\tilde\rho) ={\pi \over 2} n \, ,
\ee
where $n$ is the quantum number and $\tilde p(\tilde\rho)$ is obtained from the equation $\tilde H = \Deltax$ where $\Deltax$ is the energy of the string in units of AdS radius. The integral limit $\tilde\rho_0$ is determined so that $\tilde p(\tilde\rho_0) = 0$. This yields
\be
  \label{eq:gub}
    \int_0^{\tilde\rho_0} d\tilde\rho \left( \Deltax - {4g} \sinh\tilde\rho \right)  \textrm{sech}\, \tilde\rho ={\pi \over 2} n \,.
\ee
For small  $\tilde \rho_0$, we get the Regge trajectory for the physical mass $m$,
\be
    \label{eq:regge}
 (mL)^2 =  \Deltax^2 = {2L^2 \over \alpha'} n = 4 \pi g n\, .
\ee
On the other hand, for large $\tilde \rho_0$, eqn. \eqref{eq:gub} gives the  scaling
\be
   \label{eq:logar}
 \Deltax - n = {8g \over \pi } \log n + {\cal O}(n^0) \,.
\ee
As the authors of \cite{Callebaut:2015fsa} note, this result is very similar to that of a  rigidly rotating long string in higher-dimensional AdS \cite{Gubser:2002tv}, with the only difference being in the coefficient multiplying the logarithm.

As we will see in section \ref{sec:hammech}, there are several alternative descriptions of this simple string system. The derivation relies on integrability, for which we need to recast the system into the appropriate language. This will be the topic of the next section.

\clearpage

\section{Segmented strings   in AdS$_3$}

\label{sec:segstrings}

The folded string is somewhat complicated, because it contains both particles and string segments. In this section we consider a less singular configuration, from which the folded string can be recovered as a limiting case. First, we develop the general case and then in section~\ref{sec:foursegmented} we specialize to the system of interest: the string with four segments.

\begin{figure}[ht]
\centering
\begin{minipage}{.3\textwidth}
  \centering
  \includegraphics[width=4cm]{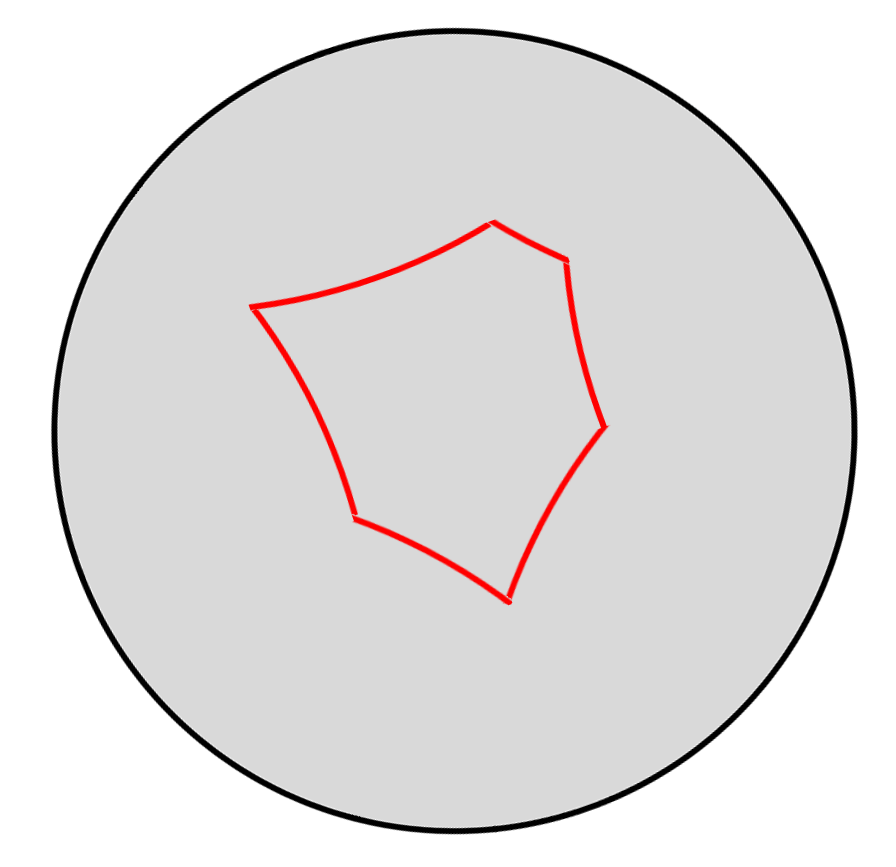}
\end{minipage}
\hfill
\begin{minipage}{.3\textwidth}
  \centering
\includegraphics[width=4.8cm]{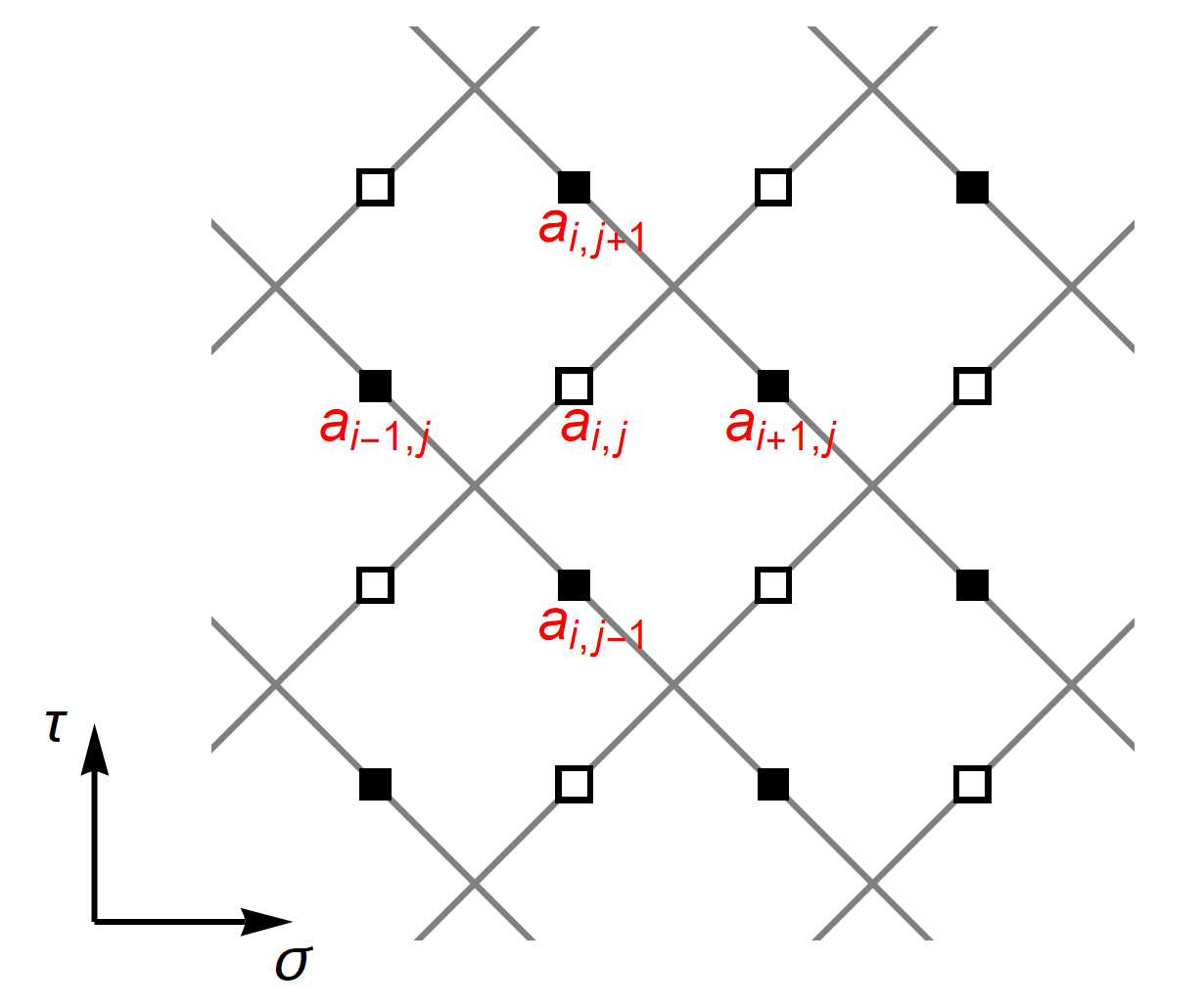}
\end{minipage}
\hfill
\begin{minipage}{.3\textwidth}
  \centering
  \includegraphics[width=3.5cm]{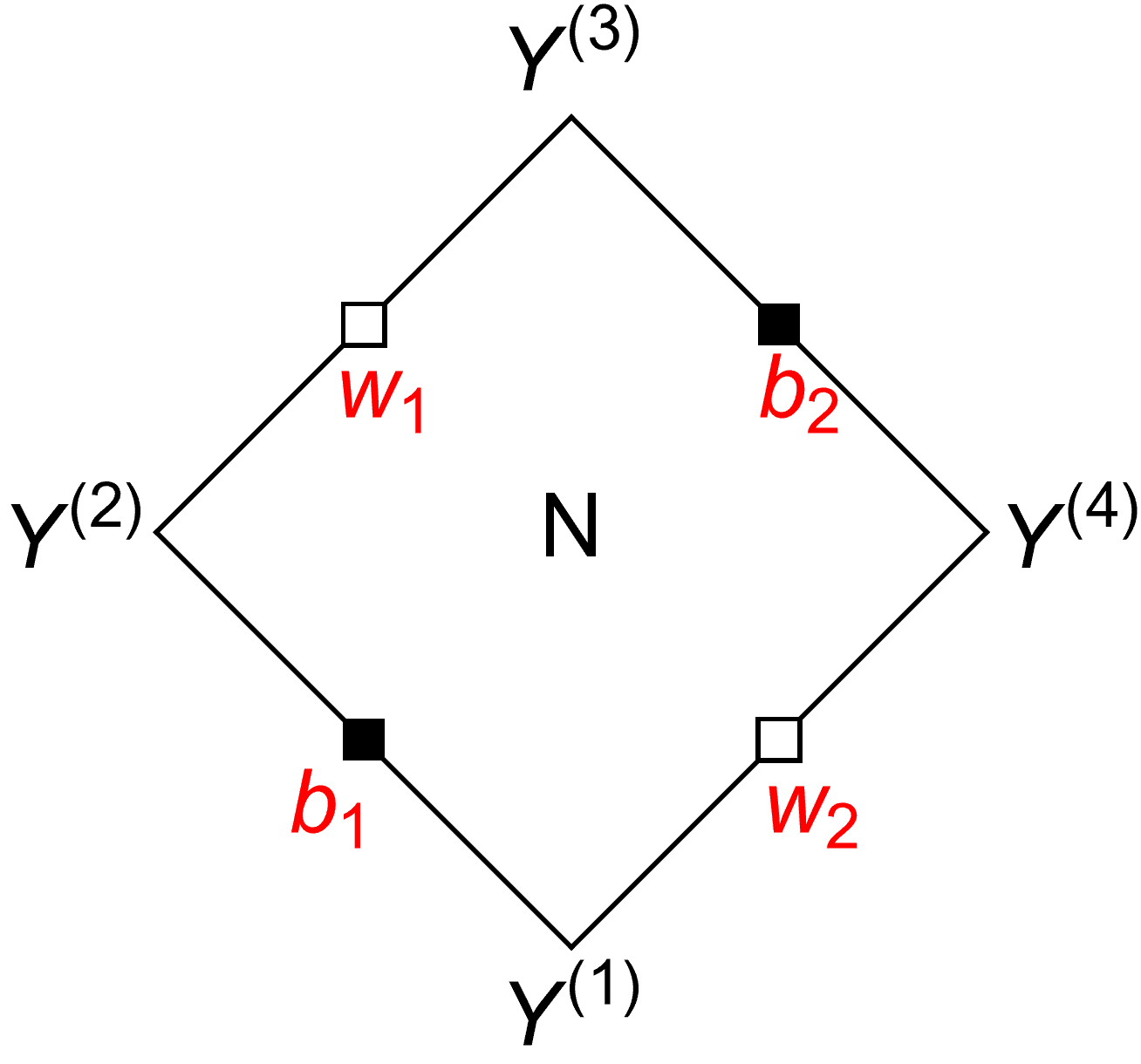}
\end{minipage}
\caption{\label{fig:sub} {\it Left:} Segmented string on a timeslice of global AdS$_3$. The example contains six kinks. {\it Middle:} A small patch of the worldsheet of a segmented string. Kinks (black lines) travel at $\pm 45^\circ$ angles, forming a  lattice on the worldsheet. The velocity of a kink is characterized by a null tangent vector  in the ambient space $\RR^{2,2}$. The corresponding $a_{ij}$ celestial variables can be written on the edges, indicated by black and white squares depending on edge orientation. Here $i$ and $j$ correspond to discretized $\sigma$ and $\tau$ coordinates, respectively. For a closed string, $a_{ij} = a_{i+n, j}$, where $n$ is the number of string segments. {\it Right:} Zooming in on a single AdS$_2$ patch on the worldsheet.
}
\end{figure}

The target space will be taken to be three-dimensional AdS space\footnote{The reason for taking AdS as a target space is (i) to make contact with the AdS/CFT correspondence \cite{Maldacena:1997re, Gubser:1998bc, Witten:1998qj} and (ii) because certain quantities (e.g. the normal vector) are more covariant in AdS.} and the classical string motion will extremize the Nambu-Goto action.
An important property of the Nambu-Goto string in AdS is that it is described by an integrable field theory \cite{Pohlmeyer:1975nb}.  Such theories can often be discretized in an exact fashion. This means that a discrete integrable model can be embedded into the field theory and discrete solutions correspond to special continuous solutions. Furthermore, the field theory can be recovered by taking a small lattice spacing limit of the discrete model\footnote{See e.g. \cite{suris} for many discrete integrable models and their continuous counterparts.}. Here we will use a particular discretization of the bosonic string---the {\it segmented string} \cite{Vegh:2015ska, Callebaut:2015fsa, Vegh:2016hwq, Gubser:2016wno, Gubser:2016zyw, Vegh:2016fcm, Vegh:2018dda, Vegh:2019any,   Vegh:2021jhl, Vegh:2021jqo}. The   embeddings are piecewise linear, i.e. the string is built from several joint segments. An example is displayed in Figure \ref{fig:sub} (left). The worldsheet of a string segment is a patch of an AdS$_2$ subspace inside AdS$_3$. Such segments are characterized by constant normal vectors. Adjacent segments are joined by {\it kinks}, which move with the speed of light. This is necessary, otherwise the string tension would immediately bend the string near the joints and the segmentedness property would be destroyed. Thus, kink worldlines form a quad lattice on the worldsheet, which is shown in Figure \ref{fig:sub} (middle).

For practical calculations we will use the split signature $\RR^{2,2}$ ambient space, in which the points of AdS$_3$ lie on a hyperboloid,
\be
  \nonumber
   Y \cdot  Y \equiv -Y_{-1}^2 - Y_0^2 + Y_1^2 + Y_2^2 = -L^2 \ , \qquad  Y \in \RR^{2,2}.
\ee
In sections \ref{sec:segstrings} and \ref{sec:curve}  we set $L=1$ for simplicity (unless otherwise stated).

As seen in  Figure \ref{fig:sub}, each AdS$_2$ patch on the worldsheet is bounded by four kink lines. If three of the vertices of the patch are known, then the fourth one can be calculated via a {\it reflection formula } \cite{Vegh:2016hwq, Gubser:2016wno}. If $Y^{(i)}$  denote the ordered vertices around the patch as in Figure \ref{fig:sub} (right), then for instance the third vertex can be computed by the formula
\be
  \label{eq:refl}
   Y^{(3)} = - Y^{(1)}  -   4 { Y^{(2)} +  Y^{(4)} \over ( Y^{(2)} +  Y^{(4)})^2 } \, .
\ee
Similarly, one can express $Y^{(1)}, Y^{(2)}, Y^{(4)}$ from the other three vertices.
If appropriate initial conditions are known, the reflection formula enables us to calculate the entire string embedding, which will extremize the Nambu-Goto action.

An alternative description uses normal vectors instead of kink vertices. Let $Y(\sigma^+, \sigma^-)$ denote the embedding function. The normal vector is defined by
\be
  \nonumber
    N_a  \propto \epsilon_{abcd} Y^b \p_+ Y^c \p_- Y^d \, \qquad \textrm{such that}  \qquad N^2 = 1 \, .
\ee
An elementary AdS$_2$ patch has a constant normal vector. Time evolution of the normal vectors on the worldsheet is governed by the expression
 \be
  \nonumber
   N^{(3)} = - N^{(1)}  +   4 { N^{(2)} +  N^{(4)} \over ( N^{(2)} +  N^{(4)})^2 } \,
\ee
where the $N^{(i)}$ are the normal vectors of four adjacent patches. This equation differs from eqn. \eqref{eq:refl} only in a sign and can also be used to compute the time evolution of the embedding  as was originally done in \cite{Vegh:2015ska}.

\subsection{Celestial variables}

In practice, the reflection formula is numerically unstable: tiny errors that inevitably arise on a computer (due to storing real numbers up to certain number of digits) get amplified over time. We will now describe a way around this problem by using embedding variables that take values on the celestial torus \cite{Vegh:2016hwq}.

Let us consider the kink `velocity' vectors in Figure \ref{fig:sub} (middle). These are defined to be the difference vectors between adjacent vertices. For instance, in Figure \ref{fig:sub} (right) the four kink velocity vectors are given by $X = Y^{(2)} - Y^{(1)}, \, Y^{(4)} - Y^{(1)}, \, Y^{(3)} - Y^{(4)}$ and $ Y^{(3)} - Y^{(2)}$, respectively. These are all null vectors, therefore by means of the spinor-helicity formalism, we can decompose them into products of (real) two-component $\RR^{2,2}$  spinors. The Nambu-Goto action (the area of the worldsheet) can be written in terms of these spinors. Due to its special form, the discrete action only depends on the ratio of the two spinor components. These ratios can be expressed using the velocity vector components $X_i$ as
\be
  \label{eq:ratios}
  {a} = {X_{-1} + X_2 \over X_0 + X_1}  \, , \qquad
  \qquad
 {\tilde a} = {X_{-1} - X_2 \over X_0 + X_1}  \, ,
\ee
for the left and right spinors, respectively. For instance, if $X =  Y^{(2)} - Y^{(1)}$, then \eqref{eq:ratios} gives $b_1$, which is shown in Figure \ref{fig:sub} (right) and $\tilde b_1$ (suppressed in the figure). These variables are dubbed {\it celestial variables}. Since they correspond to kink worldlines, the discrete field $a_{ij}\in \RR$ lives on a square lattice on the worldsheet. Here $i\in \ZZ$ and $j\in \ZZ$ correspond to discretized $\sigma$ and $\tau$ coordinates, respectively. For a closed string, $a_{ij} = a_{i+n, j}$ and  $\tilde a_{ij} = \tilde a_{i+n, j}$, where $n$ is the number of string segments. This lattice is indicated by black and white boxes in Figure \ref{fig:sub} (middle) and the coloring depends on which way the kink moves.
 Using the fact that the kink velocity vectors around a patch (with appropriate signs) must add up to zero, one can eliminate the ${\tilde a}$ variables from the action.
Finally, one obtains the equation of motion of a discrete-time  Toda lattice,
\be
  \label{eq:eom}
  \hskip -0.15cm {1\ov a_{ij} - a_{i,j+1}}+   {1\ov a_{ij} - a_{i,j-1}} =
  {1\ov a_{ij} - a_{i+1,j}}+   {1\ov a_{ij} - a_{i-1,j}} \, .
\ee
The full embedding can be recovered from the celestial variables (up to a global $SL(2)$ transformation) \cite{Vegh:2016hwq}. In what follows, it will be useful to introduce the `reflection matrix,' which depends on two adjacent celestial variables $b$ and $w$ \cite{Vegh:2019any}
\be
\label{eq:reflec}
\mathcal{R}_{b,w} =
{1\ov b-w}
\begin{pmatrix}
0 & bw+1 & bw-1 & -b-w
\cr
-1-bw & 0 & b+w & bw-1
\cr
-1+bw & b+w & 0 & -1-bw
\cr
-b-w & bw-1 & bw+1 & 0
\end{pmatrix}   \, .
\ee
Let us consider the patch on the right of  Figure \ref{fig:sub}.  The normal vector $N$ can be computed from $Y^{(2)}$ simply by letting the reflection matrix act on it,
\be
\nonumber
    N =  \mathcal{R}_{b_1,w_1}  Y^{(2)} \, .
\ee
Furthermore, from the normal vector $N$ and the celestial variables, all four $Y^{(i)}$ vertices can be computed. We have,
\be
   \nonumber
    Y^{(1)}  =  \mathcal{R}_{b_1,w_2} N   \, , \qquad
    Y^{(3)}  =  \mathcal{R}_{b_2,w_1} N   \, , \qquad
    Y^{(4)}  =  \mathcal{R}_{b_2,w_2} N   \, .
\ee
By repeatedly applying the reflection matrix on the position and normal vectors, all the other patch vertices can be computed, which finally fixes the string embedding. Note that only information from the $ a_{ij}$ variables has been used (the $\tilde a_{ij}$ variables are not independent).

\begin{figure}[h]
\begin{center}
\includegraphics[width=16cm]{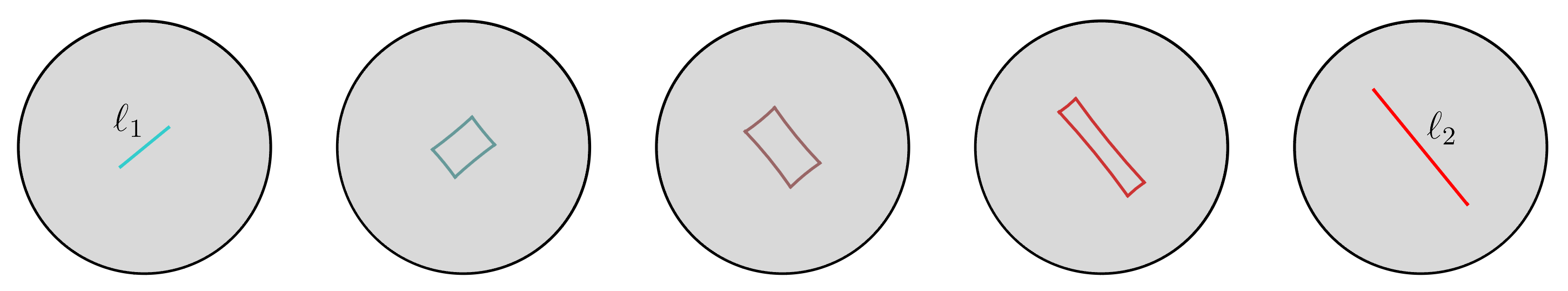}
\caption{\label{fig:phases} Snapshots of a closed string in global AdS$_3$. The motion is periodic in time. The string oscillates between the leftmost and rightmost configurations, which are characterized by their lengths $\ell_1$ and $\ell_2$.
}
\end{center}
\end{figure}

Initial conditions for the time evolution can be given by specifying two adjacent rows, e.g. $a_{i,0}$ and $a_{i,1}$ for all $i$. In the case of closed strings it is important to note  that these two rows  cannot be arbitrary real numbers. Constraints arise from the fact that the string must form a closed loop. The $a_{ij}$ variables are trivially periodic (they are chosen that way). However, the suppressed $\tilde a_{ij}$ variables must also be periodic. Since  $\tilde a_{ij}$ can be computed from $a_{ij}$ in a non-local way, this condition imposes three real constraints on $a_{ij}$. The constraints can be expressed using either the reflection matrix  \eqref{eq:reflec}, or in terms of the $2 \times 2$ Lax matrices \cite{Vegh:2021jhl}, which will be discussed in detail in section \ref{sec:curve}.

\subsection{The four-segmented string}

\label{sec:foursegmented}

In this section we discuss in detail the simplest closed segmented string, which consists of four elementary segments. The folded string from the previous section will turn out to be a special case.
Figure \ref{fig:phases} displays the motion of a generic four-segmented string in its rest frame. The snapshots are timeslices of global AdS$_3$, separated by fixed timesteps. The motion is periodic and the shape of the string gradually changes from the interval on the left (short blue line) to the interval on the right (long red line) and then back again {\it ad infinitum}. In the rest frame, the shape of any four-segmented string can be characterized by the two length scales corresponding to the intervals. In the following, we will parametrize these lengths by two constants $r_1$ and $r_2$. We will later relate these parameters to the geodesic lengths $\ell_{1}, \ell_{2}$.

\begin{figure}[h]
\begin{center}
\includegraphics[width=6.6cm]{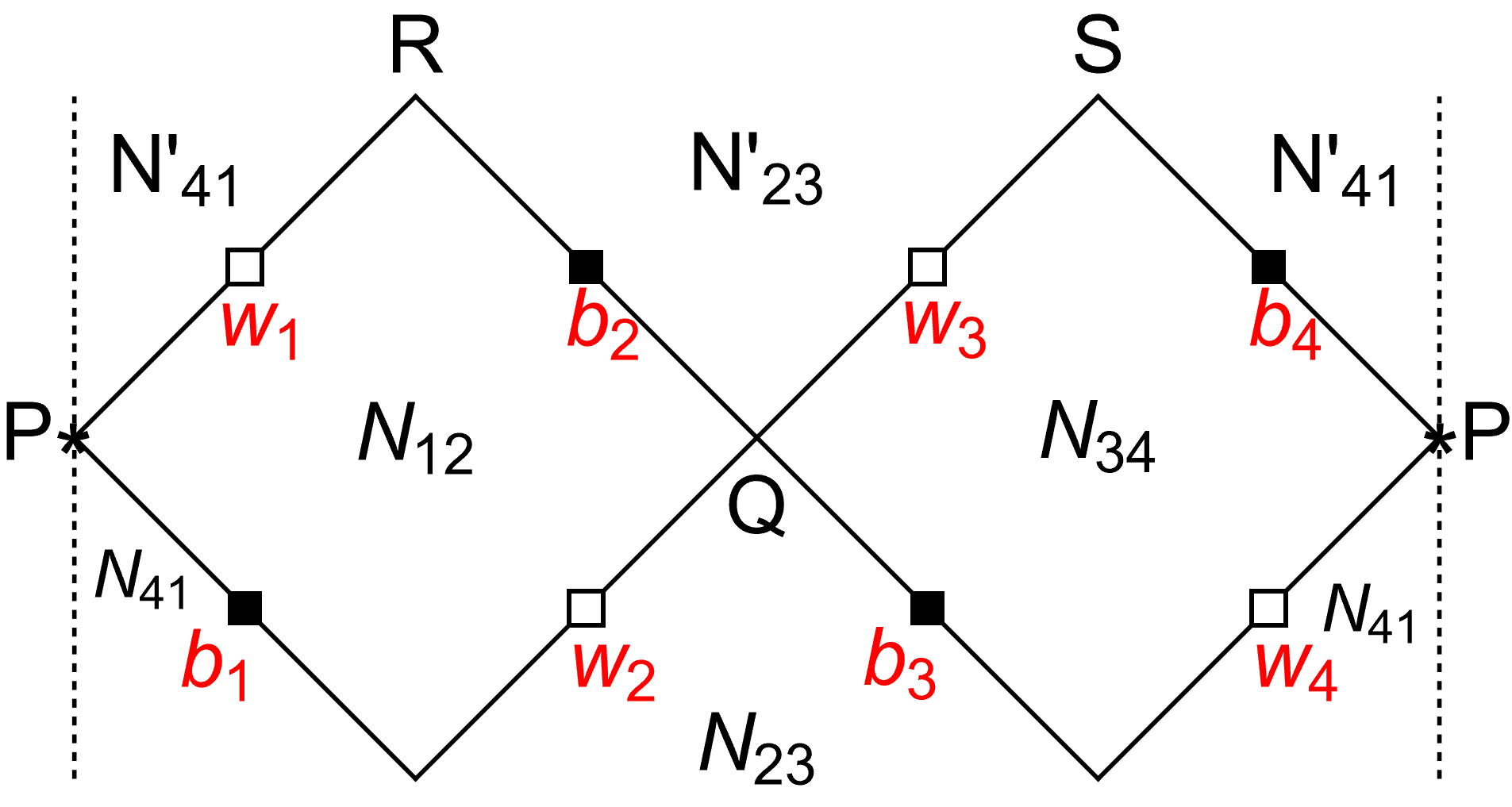}
\caption{\label{fig:4seg} Periodic worldsheet of the four-segmented string. Solid black lines indicate kink worldlines. $N_{12}, N_{23}, N'_{23}, N_{34}, N_{41}, N'_{41}$ label the normal vectors of the corresponding AdS$_2$ patches. $b_i$ and $w_j$ are celestial variables. $P, Q, R, S$ are vertices (i.e. kinks collide there).
}
\end{center}
\end{figure}

The worldsheet of the four-segmented string is shown in Figure \ref{fig:4seg}. Since the string is closed, the horizontal direction is periodic and the two dashed lines must be identified. $N_{ab}$ denote the normal vectors of the individual patches. $P$ and $Q$ denote two spacetime ``events'' where kinks on the string collide.
The symmetric segmented string with $r_1=r_2$ was described in \cite{Callebaut:2015fsa} and the corresponding normal vectors were computed in \cite{Vegh:2021jhl}. We now generalize these findings by considering the general case. We will set the AdS radius $L=1$.

Let us consider the normal vectors,
\bea
  \nonumber
  N_{12} &=& \le( r, \, -{r_2 \ov r_1}\sqrt{1+r^2}, \,  0,   \, +{d\ov r_1}   \) \, , \\
  \nonumber
  N_{23} &=& \le( r, \, + {r_1 \ov r_2}\sqrt{1+r^2},  \, -{d\ov r_2}  , \, 0 \) \, , \\
  \nonumber
  N_{34} &=& \le( r, \, -{r_2 \ov r_1}\sqrt{1+r^2}, \,  0, \, -{d\ov r_1} \) \, , \\
  \label{eq:fournormals}
  N_{41} &=& \le( r, \, + {r_1 \ov r_2}\sqrt{1+r^2},   \, +{d\ov r_2}  , \,   0 \) \, .
\eea

\noindent
where we defined the constants $r$ and $d$,
\be
  \label{eq:defrd}
  r := {\sqrt{2}r_1 r_2 \ov \sqrt{r_1^2 + r_2^2}} \, , \qquad \qquad
  d :=  \sqrt{r_1^2 + r_2^2 + 2 r_1^2 r_2^2} 
  \, .
\ee
\comment{
\bea
  \nonumber
  N_{12} &=& \le( r, \, -{r_2 \ov r_1}\sqrt{1+r^2}, \,  0,   \, +{d\ov r_1}   \) \, , \qquad
  N_{23} = \le( r, \, + {r_1 \ov r_2}\sqrt{1+r^2},  \, -{d\ov r_2}  , \, 0 \) \, , \\
  N_{34} &=& \le( r, \, -{r_2 \ov r_1}\sqrt{1+r^2}, \,  0, \, -{d\ov r_1} \) \, ,  \qquad \label{eq:fournormals}
  N_{41} = \le( r, \, + {r_1 \ov r_2}\sqrt{1+r^2},   \, +{d\ov r_2}  , \,   0 \) \, ,
\eea

\bea
  \nonumber
  N_{12} &=& \le( r, \, -{r_2 \ov r_1}\sqrt{1+r^2}, \,  0,   \, +{d\ov r_1}   \) \, , \\
  \nonumber
  N_{23} &=& \le( r, \, + {r_1 \ov r_2}\sqrt{1+r^2},  \, -{d\ov r_2}  , \, 0 \) \, , \\
  \nonumber
  N_{34} &=& \le( r, \, -{r_2 \ov r_1}\sqrt{1+r^2}, \,  0, \, -{d\ov r_1} \) \, , \\
  \label{eq:fournormals}
  N_{41} &=& \le( r, \, + {r_1 \ov r_2}\sqrt{1+r^2},   \, +{d\ov r_2}  , \,   0 \) \, .
\eea
}
It is easy to check that the normal vectors are unit vectors.
Furthermore, the scalar product of pairs of normal vectors corresponding to adjacent segments satisfy
\be
  \nonumber
  N_{12} N_{23} = N_{23} N_{34} = N_{34} N_{41} =N_{41} N_{12} = 1 \, ,
\ee
which ensures that the kinks between the segments move with the speed of light \cite{Vegh:2015ska}.

The spacetime energy of the string can be computed by a straightforward integration of the energy density as in section 4 of \cite{Callebaut:2015fsa}.
We would like to do this integral from the point $P$ to $Q$. We have $P N_{12} = P N_{41} = P N_{34} = QN_{12} = QN_{23} = QN_{34} = 0$, $P^2 = Q^2 = -1$, from which we obtain
\bea
  \nonumber
  P &=& \le( \sqrt{1+r^2}, \, {\sqrt{2}r_1^2 \ov \sqrt{r_1^2 + r_2^2}}, \,  +\sqrt{2}r_1,   \, 0   \) \, , \\
  \nonumber
  Q &=& \le( \sqrt{1+r^2}, \, {\sqrt{2}r_1^2 \ov \sqrt{r_1^2 + r_2^2}}, \,  -\sqrt{2}r_1,   \, 0   \) \, .
\eea
\comment{
\be
  \nonumber
  P = \le( \sqrt{1+r^2}, \, {\sqrt{2}r_1^2 \ov \sqrt{r_1^2 + r_2^2}}, \,  +\sqrt{2}r_1,   \, 0   \) \, , \qquad
  Q = \le( \sqrt{1+r^2}, \, {\sqrt{2}r_1^2 \ov \sqrt{r_1^2 + r_2^2}}, \,  -\sqrt{2}r_1,   \, 0   \) \, .
\ee
\bea
  \nonumber
  P &=& \le( \sqrt{1+r^2}, \, {\sqrt{2}r_1^2 \ov \sqrt{r_1^2 + r_2^2}}, \,  +\sqrt{2}r_1,   \, 0   \) \, , \\
  \nonumber
  Q &=& \le( \sqrt{1+r^2}, \, {\sqrt{2}r_1^2 \ov \sqrt{r_1^2 + r_2^2}}, \,  -\sqrt{2}r_1,   \, 0   \) \, .
\eea
}
As is well-known, the geodesic distance $\ell_1$  between $P$ and $Q$ can be computed from the dot-product,
\be
  \label{eq:geod}
   \cosh \ell_1 = -P\cdot Q = 1+4r_1^2 \, ,
\ee
which is the actual length of the string in the first plot in Figure \ref{fig:phases}. Similarly,
\be
  \label{eq:geod2}
  \cosh \ell_2 = -R\cdot S = 1+4r_2^2
\ee
gives the length of the string in the last plot.

If $\tau$ denotes global AdS time, then at the $P$ and $Q$ points $\tan \tau = {d \ov \sqrt{2} r_1^2}$ and for the $\tau-\tau$ component of the spacetime energy-momentum tensor we get (in units of  $2\pi \alpha' $),
\be
  \nonumber
  P^\tau_\tau =   {d \ov r_1} \, ,
\ee
The energy of the string is
\be
  \nonumber
  \Deltax = 2  \le| \int_{Y^1_P}^{Y^1_Q} dY^1  \,  P^\tau_\tau \ri| \, .
\ee
In order to take into account the other segment (from $Q$ back to $P$), we have multiplied the integral by two. The integral limits are the third components of the $P$ and $Q$ vectors:  $Y^1_P = \sqrt{2}r_1$ and  $Y^1_Q = -\sqrt{2}r_1$ \cite{Callebaut:2015fsa}.
Finally we get
\be
  \label{eq:energy}
  \Deltax(r_1,r_2) =
  4 \sqrt{2} d = 4 \sqrt{2}\sqrt{r_1^2 + r_2^2 + 2 r_1^2 r_2^2} \, .
\ee

\comment{
\be
  \label{eq:energy}
  \Deltax =
  mL=
  \le({L^2 \ov 2\pi \alpha' }\ri) \times 4 \sqrt{2} d =\le({L^2 \ov 2\pi \alpha' }\ri) \times  4 \sqrt{2}\sqrt{r_1^2 + r_2^2 + 2 r_1^2 r_2^2} \, .
\ee
}

If either $r_1$ or $r_2$ vanishes, then the entire motion of the four-segmented string is constrained to an AdS$_2$ subspace.
In this case, the shape of the string is just a straight line  and two of the normal vectors in (\ref{eq:fournormals})  have diverging components. The corresponding segments are vanishingly small and they move with the speed of light. In this limit they  can  be regarded  as massless particles, connected by the other two extended string segments  as shown in Figure \ref{fig:yoyo}. Finally, the celestial variables  can also be determined. Since we have a folded worldsheet, some of the kinks in Figure \ref{fig:4seg} travel on identical worldlines. Thus, the corresponding celestial variables are also equal,
\be
   \nonumber
 b_3 = w_2, \qquad  w_3 = b_2, \qquad w_4 = b_1 , \qquad  b_4 = w_1 \, .
\ee

\clearpage

\begin{figure}[h]
\begin{center}
\includegraphics[width=6cm]{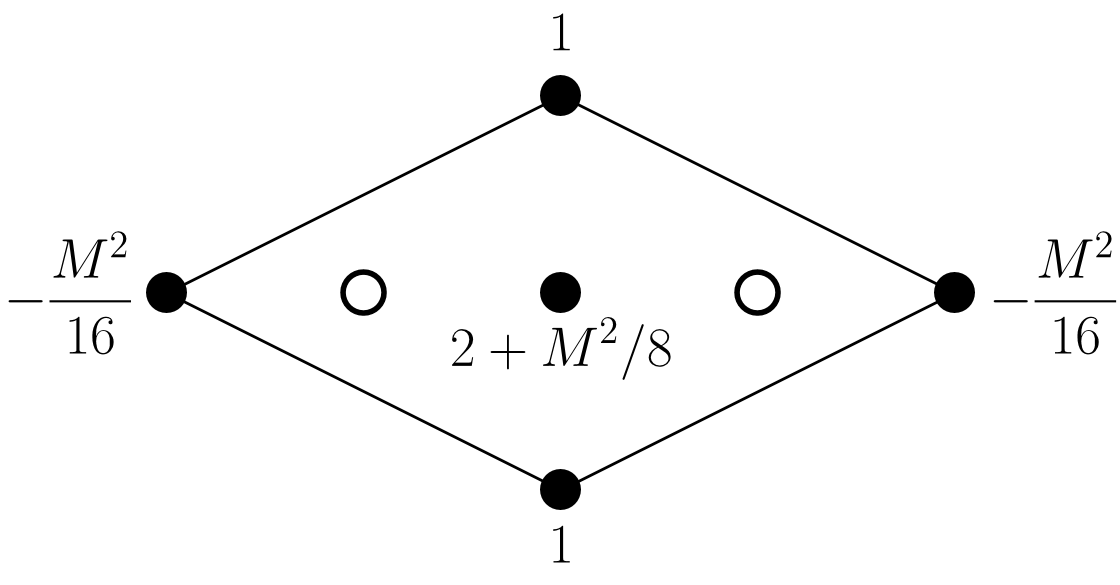}
\caption{\label{fig:toric}
Newton polygon of the spectral curve \eqref{eq:spec0} of the four-segmented string ($g=1$). All non-zero coefficients are written next to the lattice points (empty circles indicate zeroes).
}
\end{center}
\end{figure}

\section{Classical spectral curve}
\label{sec:curve}

In an integrable system the spectral curve  is an important invariant, which encodes conserved quantities.
In this section we compute the spectral curve of the four-segmented string. Although this system has only one degree of freedom (and thus it is somewhat trivial from an integrability perspective), the calculation will still be useful for it enables us to identify a natural pair of canonically conjugate variables and compute the spacetime Hamiltonian in a straightforward way, complementing the discussion in section \ref{sec:two}.

The spectral curve can be computed either by multiplying $2 \times 2$ Lax matrices, or by computing the determinant of the dressed adjacency matrix---the Kasteleyn matrix---of a relevant bipartite graph. In the following, we will discuss these methods.

\subsection{Spectral curve from celestial variables}

The elementary $2\times 2$  Lax matrix for a path starting on a kink collision vertex and ending on an adjacent AdS$_2$ patch is given by \cite{Vegh:2021jhl}
\be
  \label{eq:mono}
\Omega_{b,w}(x) =
{1 \over (b-w) \sqrt{x}}
\begin{pmatrix}
{b x - w } & {b w(1-x)  }
\cr
 { x-1  } & { b -w x  }
\end{pmatrix}
\ee
Here $x$ is a spectral parameter\footnote{Note that  \cite{Vegh:2021jhl} defines the Lax matrix in terms of the spectral parameter $\zeta \equiv \sqrt{x}$.} and $b, w$ are two celestial variables corresponding to kink worldlines which collide in a vertex. The matrix describes the parallel transport between the vertex and the patch. For instance, in Figure \ref{fig:sub} (right), $\Omega_{b_1,w_1}$ would correspond to the path from $Y^{(2)}$ to $N$. Here the symbol $N$ stands for both the patch and its normal vector\footnote{Note that there is a `duality' between AdS$_3$ and the space of normal vectors.}.

Starting at point $P$ and going around the string, the monodromy is given by
\be
  \label{eq:oms4}
  \Omega_P(x) = \Omega_{b_4, w_4}^{-1} \Omega_{b_{3}, w_{3}} \Omega_{b_2, w_2}^{-1} \Omega_{b_1, w_1} \, .
\ee
It should be mentioned that the monodromy does depend on the starting point (and changes by conjugation if we start elsewhere), but we will suppress the label as it will not be important in what follows.

In order for  the string to form a closed loop in target space, the variables must satisfy {\it closing constraints}. In terms of the $\Omega$ matrix, the constraints are given by  \cite{Vegh:2021jhl}
\be
  \label{eq:closeo}
  \Omega(x=-1) =   \mathbb{1} \, .
\ee
This equation will be referred to as the celestial closing constraint.
Note that $\Omega(x=1) =   \mathbb{1}$   is trivially true, implying that the celestial variables $a_{ij}$ are already periodic ($a_{i+4,j}=a_{ij}$). Eqn. \eqref{eq:closeo} says that the string embedding is periodic, therefore $\tilde a_{ij}$ must also be periodic.
Since $\det \Omega = 1$, \eqref{eq:closeo} constrains three out of the eight celestial variables. Three more variables can be fixed by means of an $SL(2)$ isometry transformation, which acts as a \moebius transformation on  celestial variables. The remaining two variables span the two-dimensional phase space of the string.

The spectral curve is given by
\be
  \label{eq:spec4b}
  \det\le( y \mathbb{1} + \Omega(x) \ri) = 0 \, .
\ee
The expression is quite complicated in terms of the celestial variables, especially when \eqref{eq:closeo} is used to restrict three of them. Although celestial variables do transform under $SL(2)$, the spectral curve remains invariant. Hence, the variables must appear in certain \moebiuss-invariant combinations in the equation defining the curve.

The internal energy of the string $\Deltax$ can be computed by expanding $\tr \Omega^n$ at special points, but it can also be computed directly as follows.  Figure \ref{fig:4seg} shows all the relevant celestial variables. Our goal is to determine $r_1$ and $r_2$, which enter the formula \eqref{eq:energy} for the energy.
Let us denote the components of the vertex $Q$ by
\be
  \nonumber
Q = \le( q_{-1}, \, q_{0}, \,  q_{1},   \, q_{2}   \) \, .
\ee
From $Q$, we can compute the normal vectors of neighboring patches,
\be
  \nonumber
  N_{12} =  \mathcal{R}_{b_2,w_2}  Q \, , \qquad
    N_{34} =  \mathcal{R}_{b_3,w_3}  Q \, .
\ee
Another application of the reflection matrix gives the point
\be
  \nonumber
  P =  \mathcal{R}_{b_1,w_1}  N_{12} \, .
\ee
Now using \eqref{eq:geod} and the fact that $Q^2 = -1$, we obtain
\be
  \nonumber
  r_1^2 = {(b_2 - b_1)(w_1 - w_2) \ov 2(b_1 - w_1)(b_2 - w_2)} \, .
\ee
Note that the components of $Q$ have dropped out of the final expression. In order to compute $r_2$, we proceed to compute the points
\be
  \nonumber
  R =  \mathcal{R}_{b_2,w_1}  N_{12} \, , \qquad
  S =  \mathcal{R}_{b_4,w_3}  N_{34} \, .
\ee
Using \eqref{eq:geod2} and  $Q^2 = -1$, we obtain
\be
  \nonumber
  r_2^2 = {(b_3 - b_4)(w_1 - w_2)(b_2-w_3)^2 \ov 2(b_2 - w_1)(b_2 - w_2)(b_3 - w_3)(w_3 - b_4)} \, .
\ee
This result is slightly more complicated than the one for $r_1$, because  $R$ and $S$ lie a bit farther away from the two rows of celestial variables than $P$ and $Q$.
Let us now fix  three of the variables, say $w_3, b_4, w_4$, by imposing \eqref{eq:closeo}. We get \cite{Vegh:2021jqo}
\bea
  \nonumber
  w_3 &=& {b_1 w_1(b_3-w_2)+b_2 w_2(w_1-b_3)+b_1 b_2(w_2-w_1) \ov b_1(b_3-w_1)+b_3(w_1-w_2)+b_2(w_2-b_3) }  \,  , \\
  \label{eq:csol}
  b_4 &=& {b_1 b_2(b_3+w_1-w_2) + w_1 w_2(b_3-b_2)
  + b_1 w_1(w_2-2 b_3) \ov
  b_1(b_2-b_3)- w_1 b_3 + b_2(b_3-2 w_2)+w_2(w_1+b_3)} \, , \\
  \nonumber
  w_4 &=& {w_1 b_2(b_3-w_2)+ b_1(b_2(w_1-w_2)+w_2(w_1+b_3)-2 w_1 b_3)
  \ov b_2(w_1+b_3-2 w_2) + b_1(w_2-b_3)+b_3(w_2-w_1)} \, .
\eea
Using these expressions and eqn. \eqref{eq:energy} for $\Deltax(r_1, r_2)$,  it is easy to see that the spectral curve can be rewritten in the invariant form,
\be
  \label{eq:spec0}
   y + y^{-1}   - {\Deltax^2 \ov 16}(x^2 +x^{-2}) +2 +{\Deltax^2 \ov 8}  = 0 \, .
\ee
Finally, all the explicit celestial dependence is gone from the equation. The Newton polygon of the curve is shown in Figure \ref{fig:toric}.

\clearpage

\begin{figure}[ht]
\centering
\begin{minipage}{.27\textwidth}
  \centering
  \includegraphics[width=2.8cm]{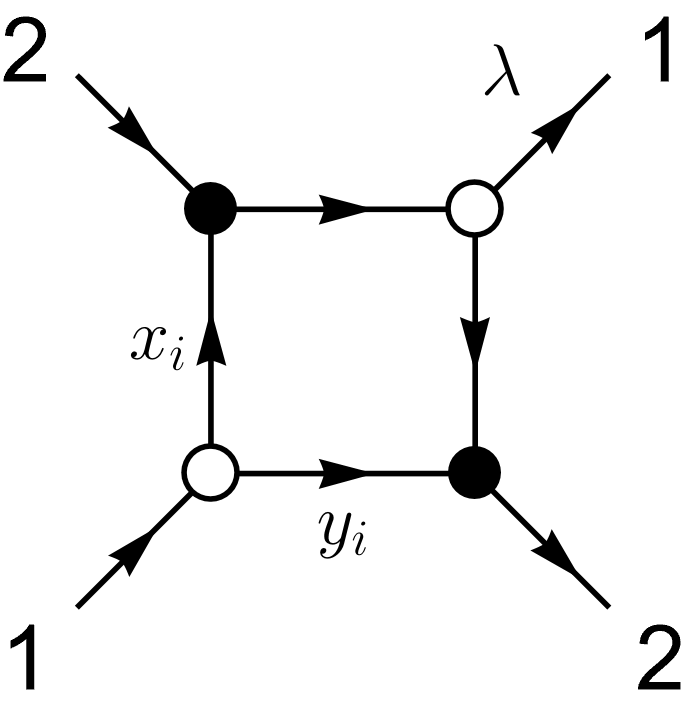}
\end{minipage}
\begin{minipage}{.33\textwidth}
  \centering
\includegraphics[width=6.5cm]{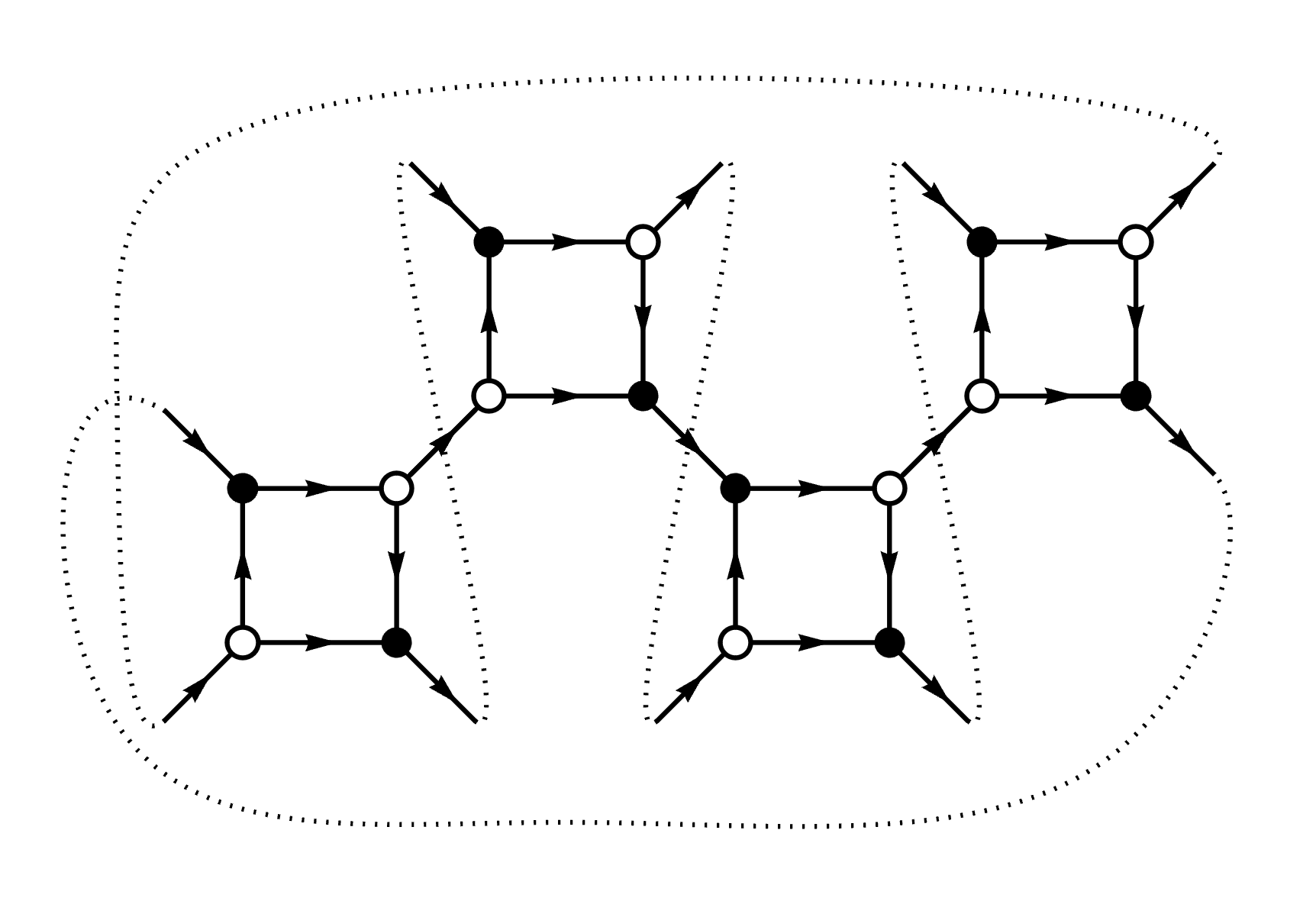}
\end{minipage}
\hfill
\begin{minipage}{.28\textwidth}
  \centering
  \includegraphics[width=5cm]{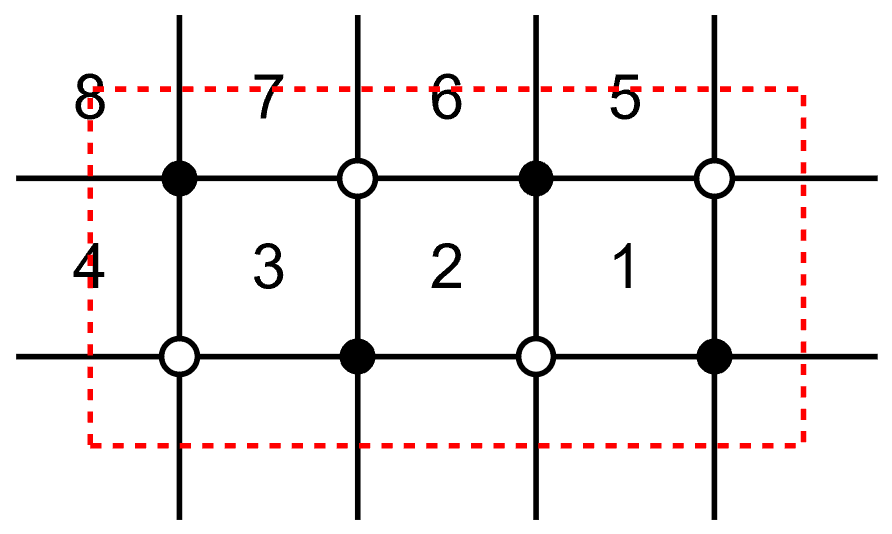}
\end{minipage}
\caption{\label{fig:lego0} {\it Left:} Elementary network (or on-shell diagram). {\it Middle:} The four-segmented string requires gluing four elementary networks in a periodic way.  {\it Right:} Collapsing nodes of the same color gives the brane tiling: a doubly periodic bipartite graph.
}
\end{figure}

\subsection{Spectral curve from tiling variables}

\label{sec:spectiling}

The spectral curve of segmented strings can also be computed using an alternative technique based on {\it on-shell diagrams} or {\it brane tilings}\footnote{The name ``brane tiling'' comes from the study of D-branes probing non-compact toric Calabi-Yau spaces. In this context, the graphs indicate brane configurations in a certain duality frame. For a non-comprehensive list of papers on the subject, see \cite{Hanany:2005ve, Franco:2005rj, Hanany:2005ss, Feng:2005gw}. For on-shell diagrams, see \cite{Arkani-Hamed:2012zlh}.}. This technique was first described by the author in \cite{Vegh:2021jqo}. The objects in question are doubly periodic planar bipartite graphs, i.e. they can be drawn on the surface of a torus without crossing edges. An example is shown in Figure~\ref{fig:lego0} (right), where the fundamental domain is indicated by a dashed rectangle.

Natural variables associated to the tiling are \moebiuss-invariant and we can relate them to celestial variables by taking cross-ratios of the latter.
We will use the notation
\be
  \nonumber
  (a, \, b; \, c, \, d) \equiv {(a-b)(c-d) \ov (a-d) (b-c)} \, .
\ee
We adopt the following definition of the tiling variables $x_{ij}$ and $y_{ij}$ \cite{GEKHTMAN2016390, Vegh:2021jqo},
\be
  \label{eq:xdef}
  x_{ij} := -(a_{i+1,j+1}, \, a_{i+2,j};  \, a_{i+1,j}, \, a_{i j} ) \, ,
\ee
\be
  \label{eq:ydef}
  x_{ij}  +y_{ij} := -(a_{i+1,j+1}, \, a_{i+2,j+1};  \, a_{i+1,j}, \, a_{i j} ) \, ,
\ee
Discrete time evolution (i.e. $j \to j+1$) is given by the formulas \cite{gekhtman2012higher}
\be
  \label{eq:xyevol}
  x_{i,j+1} = x_{i-1,j} {x_{ij} + y_{ij} \ov x_{i-1,j} + y_{i-1,j}} \, , \qquad
  y_{i,j+1} = y_{i,j} {x_{i+1,j} + y_{i+1,j} \ov x_{ij} + y_{ij}} \,   .
\ee
The expressions \eqref{eq:xdef}-\eqref{eq:ydef} contain celestial variables from rows $j$ and $j+1$. For the four-segmented string, these two rows are shown in Figure \ref{fig:4seg}. Henceforth, we will suppress the index $j$. Altogether we have defined eight tiling variables: $x_i$ and $y_i$ for $i=1\ldots 4$, but they are not independent due to the closing constraints.

The tiling can be constructed from elementary building blocks. The elementary network is depicted in Figure \ref{fig:lego0} (left)\footnote{In the context of on-shell diagrams, this is a four-particle tree-amplitude \cite{Arkani-Hamed:2012zlh}.}. It contains two black and two right nodes and directed links between them. The direction of the edges are defined such that white nodes have  two outgoing links, while black nodes only have one. We associate weights to  the edges as shown in the Figure. In addition to $x_i, y_i$, we have introduced a spectral parameter $\lambda$, which takes into account that the outgoing links are shifted on the right-hand side of the diagram.

We define the weight of a directed path as the product of its edge weights.
The ``boundary measurement matrix'' of the elementary network is a $2\times 2$   matrix. Its rows and columns are labeled by  ingoing and outgoing links and the elements are the sums of weights of all directed paths connecting the two endpoints. It is easily computed \cite{GEKHTMAN2016390}
\be
  \label{eq:mmatrix}
   B_i(\lambda) =
\begin{pmatrix}
 -\lambda x_i & x_i + y_i \\
 -\lambda & 1
\end{pmatrix} \, .
\ee
Unlike \eqref{eq:mono}, this matrix is \moebiuss-invariant. The Lax matrix is the normalized boundary measurement matrix,
\be
  \nonumber
   L_i :=  {B_i \ov \sqrt{\det B_i}} \, ,
\ee
such that  $\det  L_i = 1$.
The Lax monodromy is given by the product of Lax matrices\footnote{Note that the order of multiplication is reversed compared to \eqref{eq:oms4}.},
\be
  \label{eq:ell}
  L(\lambda) =    L_1   L_2   L_3   L_4  \, .
\ee
The spectral parameter $\lambda $ can be identified with $x$ from the previous section.
The closing constraints can be written as\footnote{The general $N$-segmented case will be discussed elsewhere.}
\be
  \label{eq:ccl}
  L(1) =    \mathbb{1} \, , \qquad  L(-1) =   \mathbb{1} \, .
\ee
These equations are \moebiuss-invariant.
An interpretation of the equations is that the underlying  celestial variables must be periodic. If for given tiling variables  $x_i$ and $y_i$ the constraints are not satisfied, then one can still locally assign celestial variables so that \eqref{eq:xdef} and \eqref{eq:ydef} are true, but these celestial variables will suffer an $SL(2)$ monodromy as we go around the string.
The requirement that the monodromy is trivial for both $a_{ij}$ and $\tilde a_{ij}$ reduces the number of independent variables by six\footnote{In the previous section, we  arrived at a two-dimensional phase space in an asymmetric way. There $\Omega(x)$ explicitly contained $a_{ij}$, which were assumed to be periodic from the start. The closing constraint \eqref{eq:closeo} eliminated three celestial variables, and a further reduction came by considering \moebiuss-invariance.}.
Solving the constraints gives
\be
  \label{eq:sols}
  x_1 = x_3 =  -X^{-1} \, , \quad x_2 = x_4 = X \, , \quad y_1 = y_3 =  X^{-1}-X-Y \, , \quad y_2 = y_4 = Y \, ,
\ee
where we parametrized the 2d phase space by $X$ and $Y$. A straightforward calculation shows that these relations follow from  \eqref{eq:xdef} and \eqref{eq:ydef} by plugging in \eqref{eq:csol}.
From \eqref{eq:xyevol} we get the discrete time step,
\be
  \label{eq:disctimestep}
  X \mapsto {1 \ov X} \, , \qquad Y \mapsto -Y \, .
\ee
Moreover, an explicit calculation reveals the following relationship between tiling variables and radii,
\be
  \label{eq:xyr1r2}
  X = -{ r_1 r_2 }{\sqrt{2}\ov d} \, , \qquad
  Y = -{ r_2 \ov r_1   }{1\ov d \sqrt{2}} \, ,
\ee
where we used the constant $d$ defined in \eqref{eq:defrd}. This result can be obtained in a straightforward (but lengthy) way by considering the normal vectors in \eqref{eq:fournormals}, then calculating the celestial variables in a frame where they are not singular, and then finally evaluating  $x_i$ and $y_i$  on the first discrete time slice. Since  tiling variables are \moebiuss-invariant, the result is true in any frame. We can express the radii,
\be
  \label{eq:radii12}
  r_1^2 = {x_2 \ov 2 y_2 } = {X \ov 2 Y} \, , \qquad
  r_2^2 =  {x_2 \ov 2 y_1 } = {X \ov 2 ( X^{-1}-X-Y)} \, ,
\ee
It is important to note that since $r_1, r_2 \geq 0$, the quantities on the RHS of both of the equations  are also non-negative.
It is easy to see that the discrete time step \eqref{eq:disctimestep} simultaneously flips the sign of both of these quantities.

For the internal energy of the string we get
\be
  \label{eq:xyenergy}
  \Deltax =    4 g\le({  X \ov Y  (1-X^2 - X Y) } \ri)^\half \, ,
\ee
where we have reinstated the dependence on the AdS radius through the constant $g$.
The discussion below \eqref{eq:radii12} implies that the internal energy is a real number. Moreover, it is invariant under discrete time steps.
Using the relations in \eqref{eq:radii12}, it is easy to show that \eqref{eq:xyenergy} is equal to $\Deltax(r_1, r_2)$ in \eqref{eq:energy} (for $g=1$).

Finally, the spectral curve of the four-segmented string is given by
\be
  \label{eq:spec4bb}
    \det\le( y \mathbb{1} + L(x) \ri) = 0 \, .
\ee
Using \eqref{eq:sols} and \eqref{eq:xyenergy}, the spectral curve can again be cast in the form of \eqref{eq:spec0}.

\clearpage

\begin{figure}[h]
\begin{center}
\includegraphics[width=10cm]{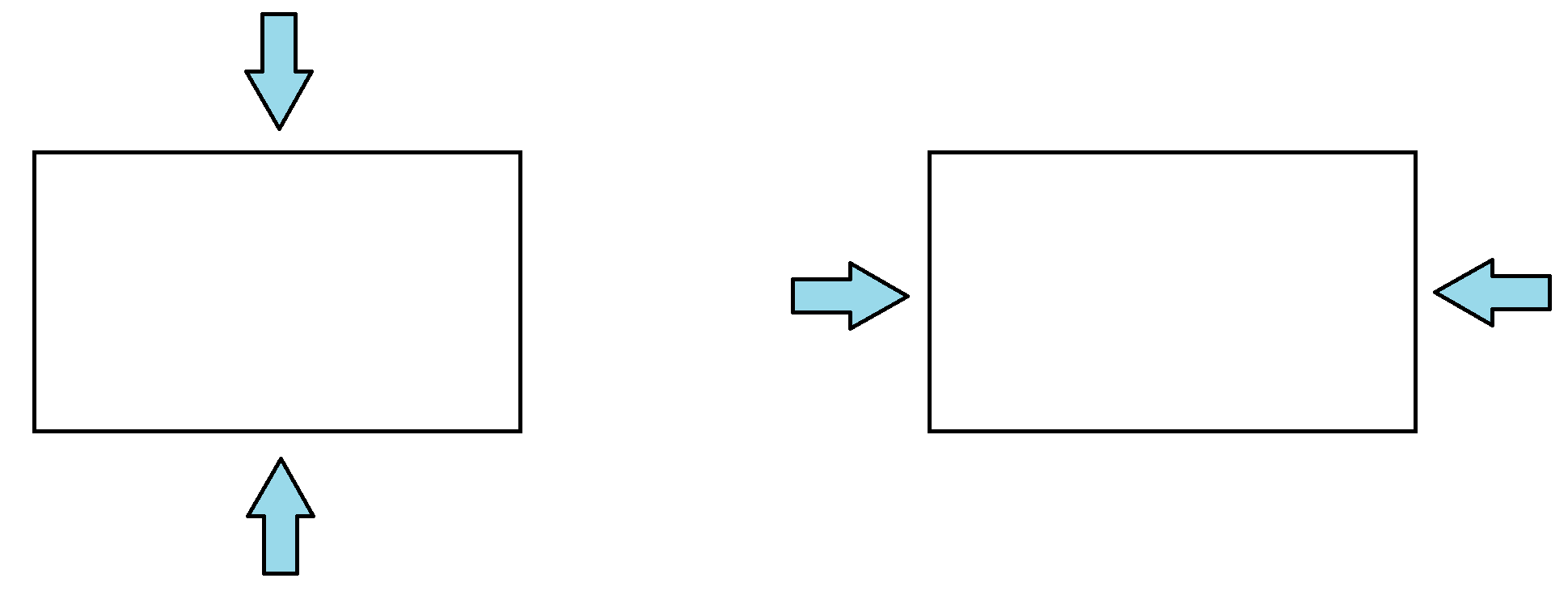}
\caption{\label{fig:compress}
The four-segmented string can be squashed into a one-dimensional folded string by sending either $r_1 \to 0$ or $r_2 \to 0$ while keeping  $\Deltax$ fixed. The spectral curve is blind to this squashing procedure.
}
\end{center}
\end{figure}

\subsection{Squashing the four-segmented string}

\label{sec:squash}

A remarkable fact is that the spectral curve does not contain complete information about the shape of the segmented string. Although the  string can be characterized in  an $SO(2,2)$-invariant way by two non-negative parameters, namely $r_1$ and $r_2$ as we discussed in section \ref{sec:foursegmented},  the spectral curve in (\ref{eq:spec}) only depends on these numbers via the energy function $\Deltax(r_1, r_2)$, which was given in \eqref{eq:energy}.
This means that we cannot unambiguously reconstruct the string embedding based purely on information contained in the spectral curve. In fact, we can squash the string embedding by sending either $r_1$ or $r_2$ all the way to zero while keeping the energy fixed. The resulting string will have the same phase space as far as the internal (\moebiuss-invariant) degrees of freedom are concerned. Moreover, since the squashed string moves in an AdS$_2$ subspace, it is nothing but the folded string, which was described in section \ref{sec:two}.

As a  consequence, we will be able to find a canonical transformation between the phase spaces of the four-segmented string  in AdS$_3$ and  the folded string  in AdS$_2$. (Note that there are two choices for this map, because as Figure \ref{fig:compress} shows, we can set either $r_1$ or $r_2$ to zero.)
Henceforth we will exploit this result and, with a slight abuse of language (and geometry), we will talk about a folded string and particles at its endpoints even when we discuss the four-segmented string.

\clearpage

\section{Hamiltonian mechanics}

\label{sec:hammech}

\subsection{Canonical variables on the spectral curve}

\label{sec:canvar}

In the previous section we have obtained the spectral curve of the four-segmented string (see Figure \ref{fig:phases}). It is given by the equation
\be
  \label{eq:spec}
   y + y^{-1}   - {\Deltax^2 \ov 16}(x^2 +x^{-2}) +2 +{\Deltax^2 \ov 8}  = 0 \, ,
\ee
where $\Deltax$ is the energy of the string, while $x$ and $y$ are spectral coordinates.
The spectral curve corresponds to both the discrete model, whose equation of motion is given in \eqref{eq:eom}, as well as the continuous-time sigma model, in which the segmented string is a somewhat singular solution \cite{Vegh:2021jhl}.

In order to make contact with  commonly used spectral parameters in the literature, let us define the Zhukovsky variable  $\chi$ (which we will only use here)
\be
  \label{eq:zhu}
  x =: {\chi - 1 \ov \chi +1} \, ,
\ee
and then immediately switch to another spectral parameter $u$ by taking a further {\it Zhukovsky transformation},
\be
  \label{eq:u}
u := (\chi + {\chi}^{-1})g  \, ,
\ee
where we have introduced the constant $g$. After a corresponding rescaling of the energy $\Deltax~\to~{\Deltax\over g}$, the spectral curve \eqref{eq:spec} becomes
\setlength\fboxsep{0.105cm}
\be
  \label{eq:specu}
 \boxed{ \quad   y + y^{-1} + 2 - {\Deltax^2 \ov u^2 -4g^2}  = 0 \, . \ \ }
\ee
Let us define the quasimomentum $p$ by setting $y = e^p $. Based on results for the smooth string\footnote{See \cite{Dorey:2006mx} and the thesis \cite{Vicedo:2008ryn} for the case where the string is embedded into $\RR \times S^3$.}, we expect that $p$ and $u$ form a pair of canonically conjugate variables: $\{ u, p \} = 1$.
They can be thought of as natural ``separated'' variables, even though the system has only one degree of freedom\footnote{For a review of separation of variables, see \cite{sklyanin95} and references therein.}.
The quasimomentum is obtained from the spectral curve
\be
  \label{eq:quasi}
  p(u) = \cosh^{-1} \le[{ \Deltax^2   \ov 2(u^2 - 4g^2) } -1 \ri]
= 2\cosh^{-1} \le[{  \Deltax   \ov 2\sqrt{u^2 - 4g^2} }  \ri] \, .
\ee
Similarly to section \ref{sec:two}, the system is quantized semiclassically by imposing the condition
\be
  \nonumber
   {\mathcal{S} \over 4} \equiv \int_{2g}^{u_0} du \,p(u)  = {\pi \over 2} n   
\ee
for the quarter period.
The upper limit of the integration is the turning point defined by $p(u_0)=0$, which gives $u_0= \sqrt{{\Deltax^2\ov 4}+4g^2}$.
The integral can be evaluated  with the result
\be
  \nonumber
 \mathcal{S} = \frac{\pi  \Deltax }{2}-2 g \log  \left(\frac{\Deltax ^2}{16}+g^2\right)+4 g \log g - \Deltax \tan^{-1} {4g \ov \Deltax} \, ,
\ee
and from this expression the results in section \ref{sec:two} can be derived.

\subsection{$\mathfrak{sl}(2)$
Ruijsenaars-Schneider  Hamiltonian}

\label{sec:hami}

Using the expression for the quasimomentum in  (\ref{eq:quasi}), we can express the spacetime Hamiltonian from $p(\Deltax, u) = p$. We get 
\setlength\fboxsep{0.15cm}
\be
  \label{eq:hamipu}
\boxed{ \quad     H(p, u) = \Deltax = 2 \cosh\le({p \ov 2}\ri) \sqrt{u^2 - 4g^2} \, . \quad }
\ee
Here $p$ and $u$ are canonically conjugate variables.
The Hamiltonian resembles that of the Ruijsenaars-Schneider (RS) model \cite{RUIJSENAARS1986370}, which is a one-parameter deformation of the Calogero-Moser model and it is relativistic in a certain sense \cite{Braden:1997nc}.
Analogously to the original RS model we can define the momentum and  boost generators,
\be
  \nonumber
   P  := 2 \sinh\le( {p \ov 2} \ri) \sqrt{u^2 - 4g^2} \, , \qquad
    B := -2u \, .
\ee
Together they form an $\mathfrak{sl}(2)$ algebra,
\be
  \nonumber
  \{ H, \, P \} = -B \, , \qquad
  \{ H, \, B \} = P \, , \qquad
  \{ P, \, B \} = H \, .
\ee
We also have
\be
  \nonumber
  -H^2 + P^2 + B^2 = 16 g^2 \, .
\ee

As an aside, we note that the above expressions for the generators naturally show up as components of an $\RR^{2,1}$ embedding of AdS$_2$, if we parametrize it using ``Schwarzschild'' coordinates,
\be
  \nonumber
  (X_{-1}, \, X_0, \, X_1 ) = (B(p,u), \, P(p,u), \, H(p,u)) \, ,
\ee
where $p$ and $u$ are identified with Schwarzschild time and radius, respectively. Time evolution generated by $H$ corresponds to rotation in the $(X_{-1}, \, X_0)$ plane, which is simply global time evolution in AdS$_2$. As global time passes, at $u = \pm 2g$ we cross the horizon, which is where $p$ diverges. This corresponds to the collision of the two particles in Figure \ref{fig:yoyo} (or the collision of kinks in the four-segmented case). The phase space is an orbifold of AdS$_2$ in which the ``black hole'' horizons of both exterior regions are identified with the ``white hole'' horizons of the other region.

\begin{figure}[h]
\begin{center}
\vskip -0.3cm
\includegraphics[width=9cm]{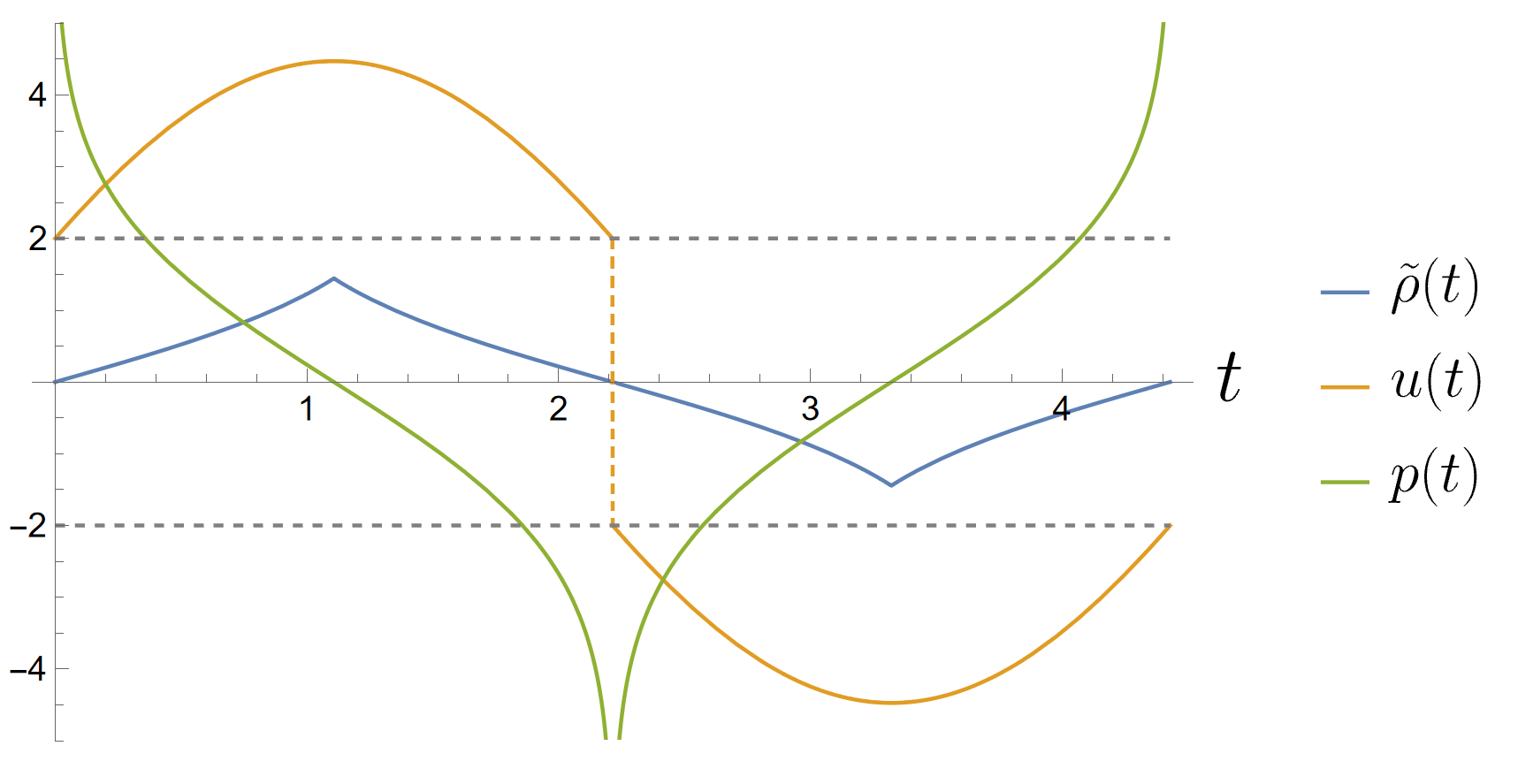}
\caption{\label{fig:motionplot}
Time evolution of dynamical variables of the folded string in AdS$_2$ (with $g=1$). Here $\tilde\rho$ is the physical position of one of the particles at the end of the string (see Figure \ref{fig:yoyo}), while $p$ and $u$ denote spectral curve coordinates on the phase space. The particles collide at $u=\pm 2g$ where $|p|\to \infty$. These two points in phase space are therefore identified.
}
\end{center}
\end{figure}

In order to understand the ``meaning'' of $p$ and $u$ and examine which regions are physically allowed, we need to find an explicit canonical transformation to physical coordinates.
Assuming $p$ and $u$ are real,
$H(p, u)$ can be transformed into the Hamiltonian  (\ref{eq:hami})  by setting $g={L^2 / ( 2 \pi \alpha')}$, $\tau = t$, and applying the following transformation (found by trial and error),
\be
  \label{eq:cantran}
  \boxed{\quad
  \tilde p(p,u)= P = 2 \sinh\le( {p \ov 2} \ri) \sqrt{u^2 - 4g^2}\, , \qquad
  \tilde\rho(p,u)  =  2  \tanh^{-1} \le(e^{-{|p|\ov 2}} \sqrt{|u|-2g \ov |u|+2g}   \ri) \, \textrm{sgn} \, u \, .
  }
\ee
Physical values of $u$ satisfy $|u|\ge 2g$.
At $u=\pm 2g$, we get $\tilde \rho=0$ and the two particles collide. The momentum $p$ diverges as seen in Figure \ref{fig:motionplot}. Since $u=\pm 2g$ correspond to the same string configuration, they should be identified.
The above transformation  is non-analytic both at these points and  at the turning points where $p$ vanishes. On the segmented string these correspond to kink collision events. It is straightforward to check that away from these one-dimensional lines in the phase space,  \eqref{eq:cantran} is in fact a canonical transformation.

Figure \ref{fig:motionplot} shows the time evolution of various dynamical variables for an yo-yoing string. Let us shift the time coordinate so that the string is pointlike at $t=0$.  Then, $\tilde\rho$ reaches its maximal value at
\be
  \nonumber
  t_\textrm{max}= \tan^{-1}{\Deltax \ov 4g}   \, .
\ee

\begin{figure}[h]
\begin{center}
\includegraphics[width=7cm]{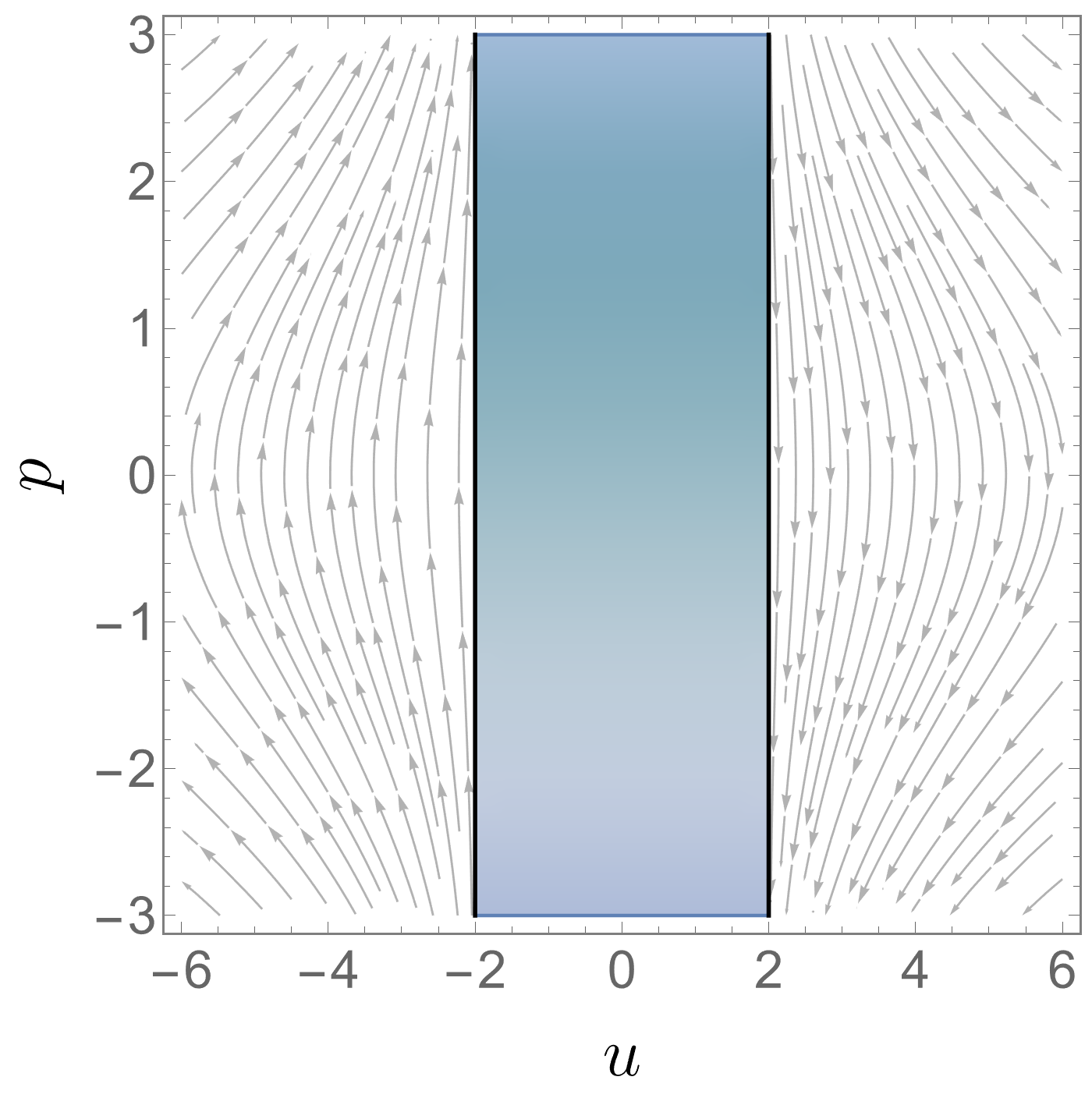} \qquad\qquad
\includegraphics[width=7cm]{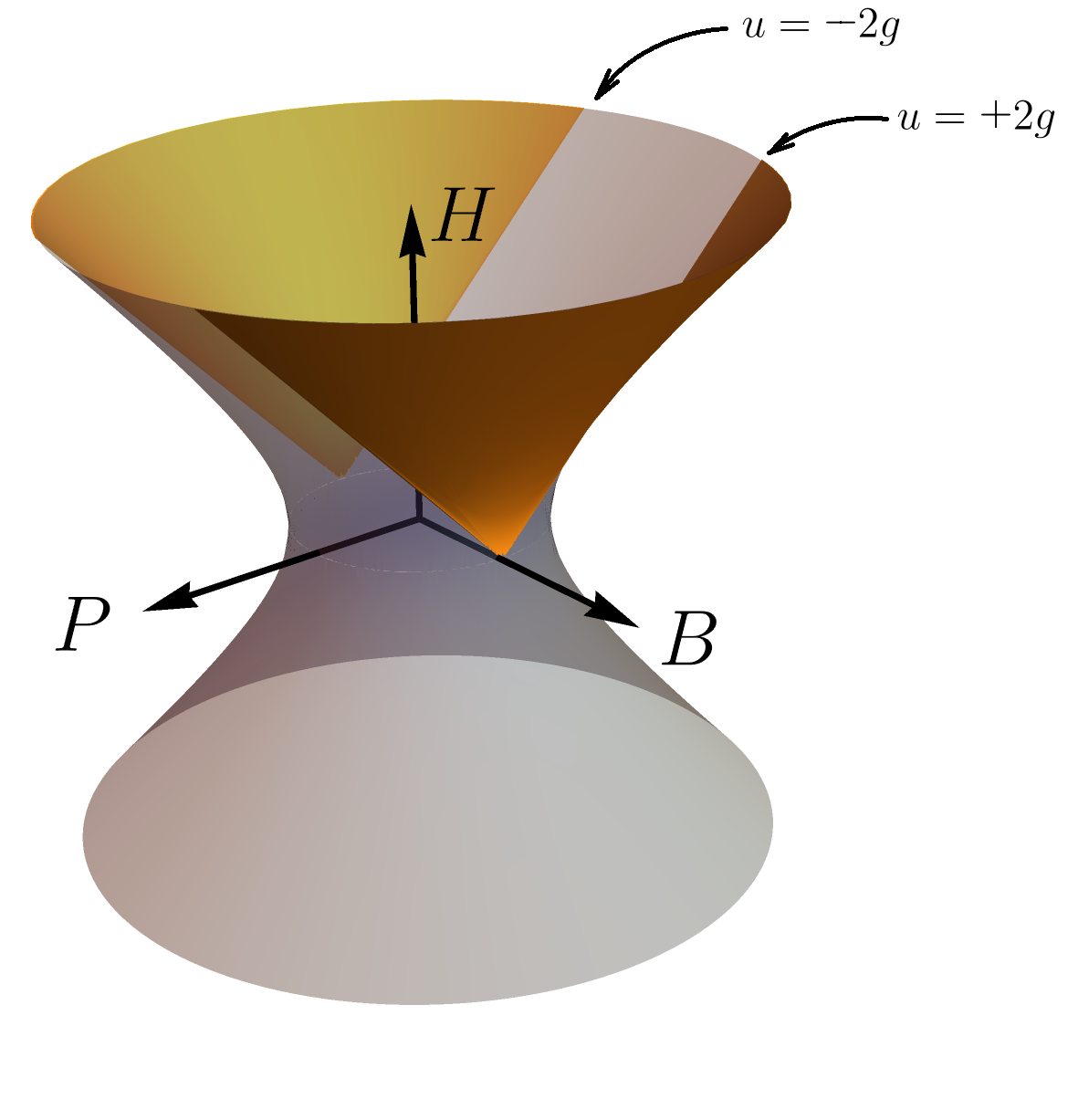}
\caption{\label{fig:flows}{\it Left:} Phase space and Hamiltonian flow of the four-segmented string in terms of natural variables $p$ and $u$ (for $g=1$). The region $|u|<2g$ is not part of the phase space.
{\it Right:} The phase space (gold regions) in $(H,P,B)$ space. The two regions should be glued together along their v-shaped boundaries at $u=\pm 2g$. The Hamiltonian is bounded from below. The Hamiltonian flow generates rigid rotation about the $H$ axis.
}
\end{center}
\end{figure}

\noindent
The period of oscillation is $4t_\textrm{max}$. We have
\bea
  \nonumber
  u(t) &=&  \begin{cases}
  2g\cos t+{\Deltax\ov 2} \sin t \qquad &(0<t<2 t_\textrm{max})\\
  -2g\cos(t-2t_\textrm{max})-{\Deltax\ov 2} \sin(t-2t_\textrm{max}) \qquad &(2 t_\textrm{max}<t<4 t_\textrm{max})
\end{cases}  \\
  \nonumber
  p(t) &=& 2 \sinh^{-1} {u'(t) \ov \sqrt{u(t)^2 - 4g^2}}  \, .
\eea
At $t=2t_\textrm{max}$, $u(t)$ jumps from $+2g$ to $-2g$ and the momentum diverges.

\begin{figure}[h]
\begin{center}
\includegraphics[width=6.2cm]{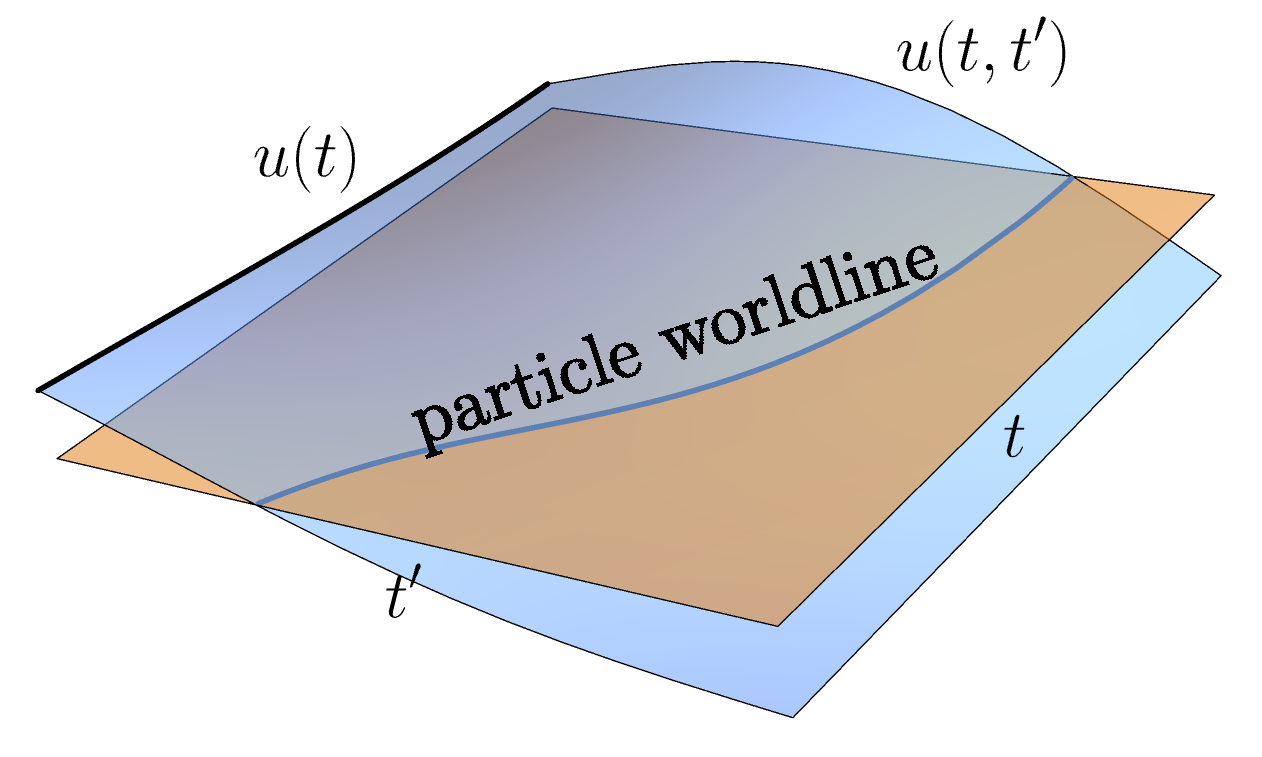}
\caption{\label{fig:levelset}
Computing soliton position from Ruijsenaars-Schneider variable $u(t)$ (schematic picture).
The flows generated by $H$ and $P$ are parametrized by $t$ and $t'$, respectively. The particle worldline $\tilde\rho(t) = -t'(t)$ is given implicitly by the level set $u(t,t') = 2g \, \textrm{sgn} \, u(t,0)$.
}
\end{center}
\end{figure}

\subsection{Physical coordinates as level sets}

Where does the transformation  \eqref{eq:cantran} come from? Here we make an observation, which might elucidate the origin of the map. Let us recall how the RS model describes the motion of solitonic particles in several (1+1)-dimensional integrable models, such as the sine-Gordon theory. First, we define functions over a two-dimensional domain by acting on the initial values with both time evolution and then the evolution generated by $P$,
\be
   \nonumber
 (p(t,t'), \, u(t,t')) :=  e^{t' P}  e^{t H} (p(0,0), \, u(0,0)) \, .
\ee
The second flow is parametrized by the $t'$ coordinate.
It should be mentioned that unlike in the original RS model, here $P$ does not commute with the Hamiltonian.

The flow generated by $P$ is given by
\be
   \nonumber
  {du \ov dt'} = \{ u, P \} =   \cosh\le( {p \ov 2} \ri) \sqrt{u^2 - 4g^2}\, , \qquad
  {dp \ov dt'} = \{ p, P \} =  - {2u \sinh\le( {p \ov 2} \ri)\ov \sqrt{u^2 - 4g^2}} \, .
\ee
RS coordinates  do not directly denote soliton positions. Instead,
soliton trajectories are defined by level sets of the $u$ function. In the case of the string $u=\pm 2g$ are special points, because they correspond to kink collisions (this is where the folded string collapses to a point). Thus, the most natural level set seems to be
\setlength\fboxsep{0.3cm}
\be
  \label{eq:levelpres}
  \boxed{ \quad u(t, t') = \pm 2g \, , \ \ }
\ee
which should be considered as an equation for $t'(t)$. We choose the sign in \eqref{eq:levelpres} so that it coincides with $\textrm{sgn} \, u(t,0)$.
Let us now consider the phase space point $(p(t,0), \, u(t,0))$ at a give time. The flow starting from this point and generated by $P$ can be integrated. For the $u$ component we get
\be
   \nonumber
   u(t,t') = u(t,0) \cosh t' + \cosh\le({p(t,0) \ov 2}\ri)\sqrt{u(t,0)^2 - 4g^2 } \sinh t' \, .
\ee
Along this flow, $P$ is constant. Now we can compute the flow time $t'$ that is needed to satisfy the level set condition  \eqref{eq:levelpres}. Note that since ${du \ov dt'}\ge 0$ always, for $u>2g$ ($u<-2g$) we will have to evolve backwards (forwards) in $t'$. The final result is precisely  (\ref{eq:cantran}) with $\tilde \rho(t) = -t'(t)$.

In summary, the transformation from spectral curve coordinates $(p,u)$  to physical coordinates $(\tilde p, \tilde\rho)$ is similar to the map that connects RS particle positions to the positions of solitons in an associated two-dimensional field theory.

\clearpage

\begin{figure}[h]
\begin{center}
\includegraphics[width=7.2cm]{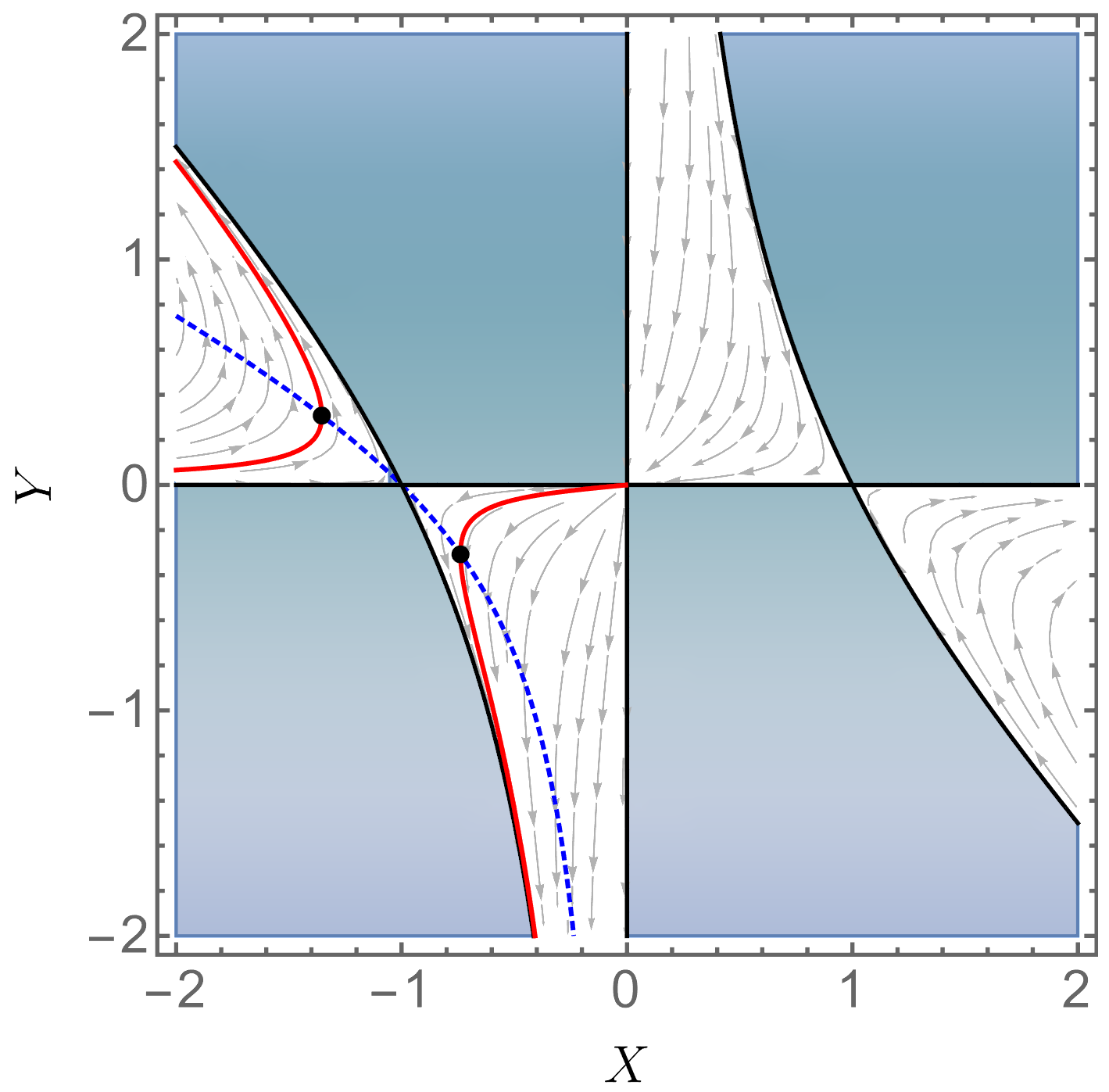} \qquad
\includegraphics[width=7.4cm]{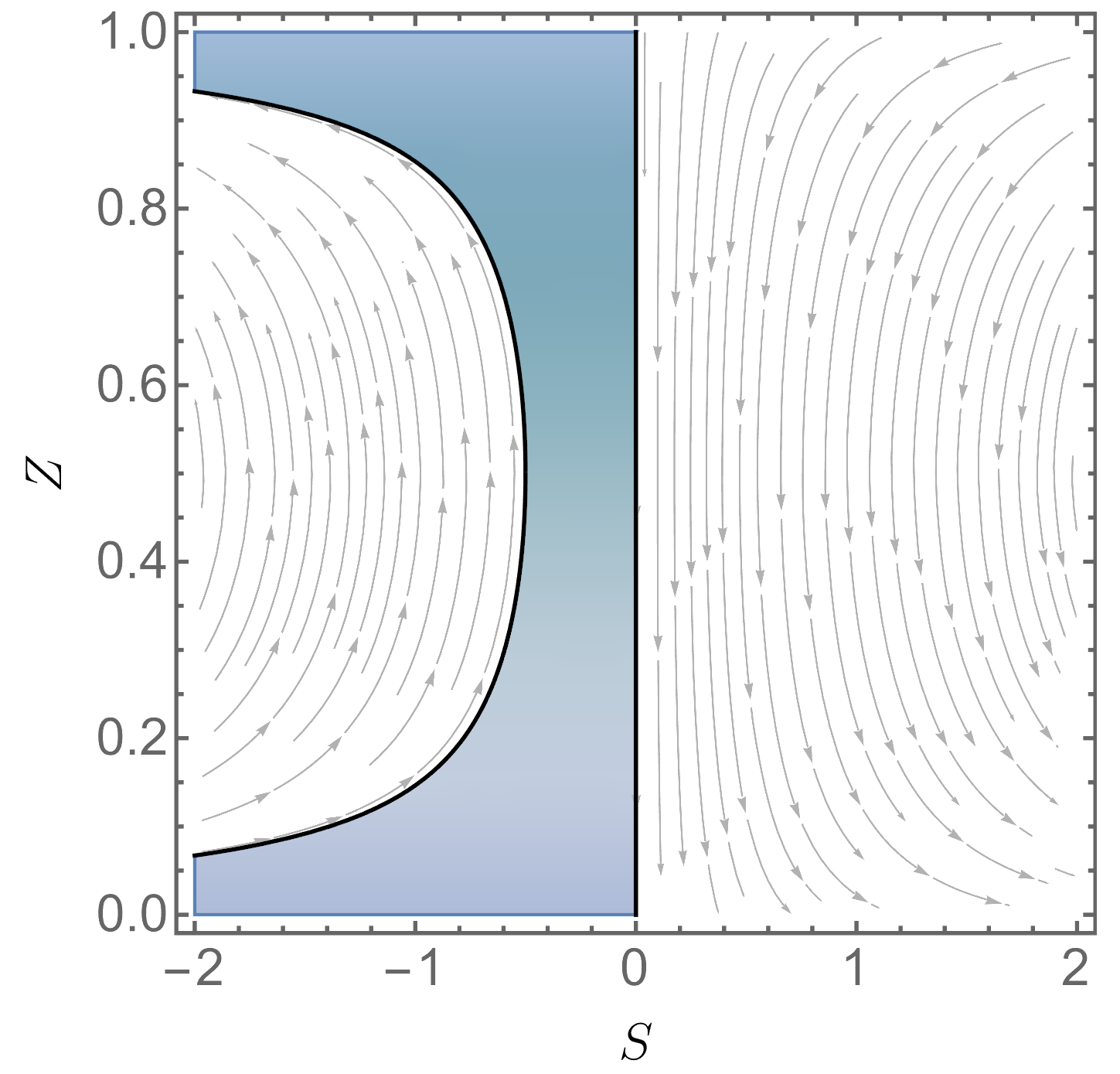}
\caption{\label{fig:flowx}
{\it Left:} Phase space and Hamiltonian flow of the four-segmented string in terms of  $X$ and $Y$ (with $g=1$). Excluded regions where  $\Deltax^2 <0$ are shown in blue. The variable $X$ is always negative and the sign of $Y$ is indefinite.
In the continuum model, the state of the system evolves according to the stream plot.
The trajectory of a particular solution is shown as a solid red line. The discrete model is embedded into the continuum theory and its dynamical variables only take specific values. For instance, in the case of $r_1=r_2$ the allowed values are indicated by a blue dotted line and the system oscillates between two points (shown by two black dots for the example) connected by \eqref{eq:disctimestep}. {\it Right:}  Phase space and Hamiltonian flow of the four-segmented string in terms of  $Z$ and $S$ of the ``first map'' (with~$g = {1\ov 32}$). The ``second map'' has a similar, mirrored diagram.
}
\end{center}
\end{figure}

\subsection{Tiling variables vs. spectral curve coordinates}

\label{sec:poisson}

Now that we have found a canonical transformation between the spectral curve coordinates $(p,u)$ and the physical coordinates $(\tilde p, \tilde \rho)$, let us proceed and find a map that connects the tiling coordinates  $(X,Y)$ to the other two pairs. Finding such a map is crucial, because it will let us calculate the Poisson brackets of $(X,Y)$. The Poisson structure together with the Hamiltonian \eqref{eq:xyenergy} will allow us to reformulate the dynamics in terms of tilings variables.

Note that up until now the tiling variables have described geometrical properties of the segmented string via \eqref{eq:xdef}-\eqref{eq:ydef}. In the following, they will be understood to be dynamical variables of the continuum time model $X(t)$ and $Y(t)$.
The change of variables that we are seeking can be obtained from the Baker-Akhiezer function and the ``magical prescription'' for the separation of variables \cite{sklyanin95}. We obtain two distinct maps as follows.

\subsubsection*{First map}

Let us consider the $2\times 2$ Lax monodromy matrix  \eqref{eq:ell} and plug in the solution of the closing constraint \eqref{eq:sols}. From $L_{12}(x) = 0$ we obtain the identification $X=x$. Then from $L_{22}(x)=e^{-p}$  we obtain two solutions for $Y$ and pick the one that gives $H(X,Y) = H(p,u)$ (see \eqref{eq:xyenergy} and \eqref{eq:hamipu}). For real coordinate values we get
\be
  \label{eq:xypu}
  X(p,u)  = x= \sqrt{u-2g \ov u+2g} \, , \qquad
  Y(p,u) =  2 g \, {\tanh\le({p\ov 2}\ri) +1 \ov \sqrt{u^2 - 4g^2}} \, \textrm{sgn} \, u \, ,
\ee
where the positive square root must be taken.
The inverse transformations are given by
\be
   \label{eq:puxytra}
   u(X,Y)=2g{1+X^2 \ov 1-X^2} \, , \qquad
   p(X,Y) = -2 \tanh^{-1}\le({1-X^2-2X Y \ov 1-X^2} \ri) \, ,
\ee
and they are invariant under the  reflection $(X,Y) \to (-X, -Y)$.

Figure \ref{fig:flowx} (left) shows the two-dimensional phase space spanned by $X$ and $Y$. The way the discrete-time model is embedded into the continuous-time model depends on  $r_1$ and $r_2$ separately. For instance, if we plug $r_1=r_2$ into \eqref{eq:xyr1r2}, then we get the relationship $2XY+X^2=1$. This curve is indicated by a dashed blue line in the Figure. The discrete model undergoes discrete time steps according to \eqref{eq:disctimestep} and the state of the system oscillates between two points. In contrast, the continuous-time model carves out a continuous trajectory (an example is shown in red).

The map between the canonical variables $(p, u)$ and the tiling variables enables us to compute
the  Poisson brackets
\be
 \label{eq:quarticpoi}
 8 g \{ X, Y \} =  {Y \ov X} - 2 XY -Y^2 + X^3 Y + X^2 Y^2 \, .
\ee
These brackets look pretty complicated\footnote{Quartic Poisson brackets also showed up in a related context in \cite{Aniceto:2008pc}. In contrast, the (quantum) integrable model of Goncharov and Kenyon  \cite{Goncharov2013} possessed much simpler quadratic Poisson brackets.} as they are quartic in $X$ and $Y$. If we want to use tiling variables to describe the Hamiltonian dynamics of the string, then simpler brackets would strongly be preferable, especially since we eventually want to quantize the system. We thus consider another
pair of coordinates  defined by\footnote{These new coordinates can be defined for more complex segmented strings by taking ratios of $x_i$ and $y_i$. They were labelled by $p_i$ and $q_i$ in \cite{GEKHTMAN2016390, Vegh:2021jqo} and correspond to the faces of the tiling.}
 \be
  \nonumber
  \boxed{ \quad  A :=  {x_2 \ov y_2 }=  {X \ov Y} \, , \qquad
  B :=  {x_2 \ov y_1 } = {X \ov   X^{-1}-X-Y } \, . \ \ }
\ee
In the discrete model at the initial time or after an even number of time steps   \eqref{eq:radii12} gives \be
  \label{eq:abeven}
  A_\textrm{even} =  2r_1^2 \geq 0 \, , \qquad
  B_\textrm{even} =  2r_2^2 \geq 0 \, .
\ee
As far as tiling variables $x_i, y_i$ with a fixed index $i$ are concerned, each time step exchanges the role of the target space and the space of normal vectors. Namely, in Figure~\ref{fig:4seg} a forward time step replaces $N_{12} \to R$, $Q \to N'_{23}$, etc., which changes the ``meaning'' of the cross-ratios of celestial variables. Accordingly, after an odd number of time steps we have a different expression in terms of the radii,
\be
 \label{eq:abodd}
  A_\textrm{odd} =  -1-2r_1^2 - {r_1^2 \ov r_2^2} <0 \, , \qquad
  B_\textrm{odd} =   -1-2r_2^2 - {r_2^2 \ov r_1^2} <0   \, .
\ee
Although the signs change, we always have $\textrm{sgn} \, A = \textrm{sgn} \, B$ (assuming they are both non-zero).
In the continuum model this continues to be true. In Figure~\ref{fig:flowx}, the boundaries separating positive and negative regions are indicated by thick black lines: the $X=0$ and the $Y=0$ lines separate positive and negative $A$ regions, while $B$ changes sign at the curved black lines.
The Poisson brackets are rather simple,
\be
 \label{eq:linpoi}
 4 g \{ A, B \} =  A+B \, .
\ee
The Hamiltonian takes the form (cf. eqn. \eqref{eq:energy})
\be
  \nonumber
  H(A,B) = 4g\sqrt{A+B+AB} \, .
\ee
We can further simplify the brackets by performing an additional change of coordinates,
\be
  \label{eq:zsdefi}
 \boxed{ \quad  Z := {B \ov A+B} \, , \qquad
  S := {4g(A+B) } \, . \ \ }
\ee
At this point, the interpretation of these new variables is somewhat obscure.
We will see in section~\ref{sec:qsc} that they provide an AdS generalization of the coordinates that appear in the 't~Hooft equation.
Eqns. \eqref{eq:abeven} and \eqref{eq:abodd} imply that the boundaries of the physically allowed region are
 \be
  \label{eq:domain1}
   0 \leq Z \leq 1  \, \qquad \textrm{and} \qquad
  S \leq -{4g  \ov Z(1-Z)}
  \ \ \textrm{or } \ S \geq 0 \, ,
\ee
which is displayed in Figure \ref{fig:flowx} (right). The two disjoint regions of the domain of $S$ correspond to the two half-periods of oscillation ($\tilde\rho > 0 $ and $\tilde\rho <0 $). It is interesting that the half-periods are treated asymmetrically in this coordinate system.
The variables satisfy  canonical brackets $\{ S, Z \} = 1 $,
and the Hamiltonian is given by
\be
 \label{eq:hamsz}
  H(S,Z) =  \sqrt{4gS+ Z(1-Z)S^2} \, .
\ee
Finally, from \eqref{eq:puxytra} we get the canonical transformation $(Z,S) \mapsto (p,u)$
\be
  \label{eq:pqzy0}
 \boxed{ \quad    p = \log{Z \ov 1-Z} \, , \qquad u =  2g+ SZ(1-Z) \, .  \ \ }
\ee

\subsubsection*{Second map}

We can also start from the other off-diagonal component of the Lax matrix and consider $L_{21}(x) = 0$ along with the corresponding $L_{11}(x)=e^{-p}$. These equations  give
\be
  \label{eq:xypu2}
  X(p,u)  = x^{-1} = \sqrt{u+2g \ov u-2g} \, , \qquad
  Y(p,u) =  -2 g \, {\tanh\le({p\ov 2}\ri) +1 \ov \sqrt{u^2 - 4g^2}} \, \textrm{sgn} \, u \, .
\ee

These are the same formulas as \eqref{eq:xypu}, except that  $X$ and $Y$ on the LHS have been transformed according to the discrete time evolution map \eqref{eq:disctimestep}.
The inverse transformations are given by
\be
   \label{eq:puxytra2}
   u(X,Y)=-2g{1+X^2 \ov 1-X^2} \, , \qquad
   p(X,Y) = -2 \tanh^{-1}\le({1-X^2-2X Y \ov 1-X^2} \ri) \, ,
\ee
which are the same as \eqref{eq:puxytra} with $u \mapsto -u$. The Poisson brackets of $X$ and $Y$ satisfy \eqref{eq:quarticpoi} if we multiply the RHS by $-1$.
In order to maintain \eqref{eq:linpoi}, we flip the signs in the definitions of $A$ and $B$,
 \be
  \nonumber
  A :=  -{x_2 \ov y_2 }=  -{X \ov Y} \, , \qquad
  B :=  -{x_2 \ov y_1 } = -{X \ov   X^{-1}-X-Y } \, .
\ee
Let us define $Z$ and $S$ as in \eqref{eq:zsdefi},
\be
  \nonumber
  \quad  Z := {B \ov A+B} \, , \qquad
  S := {4g(A+B) } \, .
\ee
The boundaries of the physically allowed region are given by
 \be
  \label{eq:domain2}
    0 \leq Z \leq 1  \, \qquad \textrm{and} \qquad
  S \leq 0
  \ \ \textrm{or } \ S \geq {4g  \ov Z(1-Z)} \, .
\ee
$Z$ and $S$ are still canonical variables and the Hamiltonian can be expressed as
\be
  \nonumber
  H(S,Z) =  \sqrt{-4gS+ Z(1-Z)S^2} \, .
\ee
Finally, for the second map we obtain
\be
  \label{eq:pqzy00}
 \boxed{ \quad  p = \log{Z \ov 1-Z} \, , \qquad u =  -2g+ SZ(1-Z) \, . \ \ }
\ee
This map is almost the same as \eqref{eq:pqzy0} except for a $4g$ shift in $u$. Furthermore, note that the domain of $S$  is also different in the two cases.

\clearpage

\renewcommand{\arraystretch}{1.2}
 \begin{table}[h!]
\begin{center}
\begin{tabular}[t]{|c|c|l|}
\hline
 {\bf coordinates} &  {\bf description} & \ \qquad\quad {\bf Hamiltonian} \\   \hline   \hline
 ($\tilde p, \tilde \rho$) & {\bf  physical center-of-mass coordinates } &  $H_\textrm{physical}=|\tilde{p}| \cosh \tilde \rho + 4g \sinh |\tilde \rho |$  \\
   &  of the folded string  (section \ref{sec:two}) &    \\ [1ex]
   ($p, u$) & {\bf spectral curve coordinates } (section \ref{sec:canvar})  & $H_\textrm{RS}=2 \cosh\le({p \ov 2}\ri) \sqrt{u^2 - 4g^2} $ \\ [1ex]
   ($Z, S$)  & {\bf  canonical tiling variables:} generalized &   $H_\textrm{ZS}= \sqrt{\pm 4gS+ Z(1-Z)S^2} $
 \\
   &  momentum fraction and action variables  &
 \\ [1ex]
 \hline
 \rule{0pt}{4ex}
 ($X, Y$) & non-canonical tiling variables (section \ref{sec:spectiling}) &  $H_\textrm{XY}= 4g \sqrt{{   X \ov Y  (1-X^2 - X Y) }  }$  \\ [1ex]
 ($A, B$) & non-canonical ``momentum'' variables  &  $H_\textrm{AB}= 4g\sqrt{\pm A \pm B+AB}$  \\ [1ex]
 \hline
\end{tabular}
\caption{Summary of various coordinates on the two-dimensional internal phase space of the four-segmented string in AdS$_3$ (or the folded string in AdS$_2$). In the  expressions for the Hamiltonian, plus and minus signs correspond to the first and the second map, respectively.
}
\end{center}
\end{table}

\subsubsection*{Mixed map}

So far we have found two distinct maps between the canonical variables $(p,u)$, which are associated to the spectral curve, and the (also canonical) tiling variables denoted by $(Z,S)$. The two maps can be transformed into one another by a shift
\be
  \nonumber
  S \mapsto S \pm {4g \ov Z(1-Z)} \, .
\ee
The physical domain in either of the $(Z,S)$ coordinate systems is somewhat awkward; see eqns. \eqref{eq:domain1} and \eqref{eq:domain2}.
Moreover, it is easy to show that in case of the first map we have the scaling,
\be
  \nonumber
  S \propto \begin{cases}
  \ {1\ov g} \qquad & \textrm{for} \, u>2g \, , \\
  \ g \qquad & \textrm{for} \, u<-2g \, ,
\end{cases}
\ee
while the second map has the opposite behavior ($g \leftrightarrow g^{-1}$). As a consequence, taking the $ g\to \infty$ flat space limit using either of these maps is cumbersome, because $S$ scales differently in the two half-periods.

In section \ref{sec:flat} we will see how a certain combination of these two maps appears in the lightcone quantization of a folded string. Namely, for $|u| \geq 2g$ (i.e. the physical region) we define the following {\it mixed map}: for  $u\geq 2g$ it is given by the first map, and for $u\leq -2g$ we switch to the second map. This drastically simplifies the boundaries of the physical region,
\be
 \nonumber
  \boxed{ \quad   0 \leq Z \leq 1  \, \qquad \textrm{and} \qquad
  S \in \RR \, . \ \ }
\ee
In the two regions, $\textrm{sgn} \, S = \textrm{sgn} \, u$, and thus the Hamiltonian can be written as
\setlength\fboxsep{0.35cm}
\be
  \label{eq:abshami}
  \boxed{ \quad H(S,Z) =  \sqrt{4g|S|+ Z(1-Z)S^2} \, .  \ \ }
\ee
This expression is one of the main results of the paper.

Taking the flat space limit is now simple. In both regions we have $S \propto g^{-1}$, which  implies that $Z(1-Z)S^2$ in  \eqref{eq:abshami} is always subleading as $ g\to \infty$. The Hamiltonian becomes $H_\textrm{flat} = \sqrt{4g|S|}$, which will also be computed in a more direct way in section \ref{sec:thooft}.

\clearpage

\section{Quantum spectral curve }

\label{sec:qsc}

\subsection{The 't Hooft equation }

\label{sec:thooft}

The celebrated  't Hooft model \cite{tHooft:1974pnl} is an $SU(N_c)$ gauge theory in (1+1)-dimensions with Dirac fermions (quarks) in the fundamental representation.
The theory is believed to share certain properties with QCD, such as quark confinement. 't Hooft solved the model in the lightcone gauge in the large-$N_c$ limit where $\lambda~\equiv~g_\textrm{YM}^2 N_c$ is kept fixed \cite{HOOFT1974461}. Here $g_\textrm{YM}$ denotes the coupling constant of the gauge theory. It turns out that free quarks are not present in the spectrum. Instead, it contains a tower of mesons, which are bound states of a quark and an antiquark. Meson wavefunctions are described by the 't Hooft equation,
\be
  \label{eq:thooft}
  \mu^2  \varphi(z) = \le[ {\alpha_1 \ov z} + {\alpha_2 \ov 1-z} \ri] \varphi(z) - \dashint_{0}^1 dz'  {\varphi(z') \ov (z' - z)^2} \, .
\ee
Here $z\in (0,1)$ denotes the  fraction of the total momentum carried by the quark, $\varphi(z) $ is the wavefunction and $\alpha_{1,2}$ are squares of renormalized quark masses. The dash on the integral sign indicates that the principal value must be taken. Assuming  appropriate boundary c  onditions for the wavefunction, \eqref{eq:thooft} is an eigenvalue equation for $\mu$, which is the meson mass in units of $\sqrt{\lambda/ \pi}$. The meson spectrum is discrete and exhibits a Regge-like behavior. Although exact solutions are not known in the general case, systematic analytic expansions do exist; see for instance \cite{Fateev:2009jf}. For a simple numerical method for calculating the spectrum, we refer the reader to \cite{Brower:1978wm} where the integral equation has been recast into a matrix eigenvalue equation. (We will use the same technique to compute the spectrum of a modified  't Hooft equation in section \ref{sec:numerics}.)
For an extensive discussion of the 't~Hooft model, we refer the reader to Coleman's lectures \cite{coleman_1985}.

In the rest of the paper we will not be concerned with 2d $SU(N_c)$ gauge theory and the original derivation of the 't Hooft  equation. We will take a different perpsective and consider~\eqref{eq:thooft} as a \schr equation, which quantizes the folded string in flat space (see Figure~\ref{fig:yoyo}). The derivation goes as follows. One writes the two-body Hamiltonian for the string with two endpoints
\be
  \nonumber
  H_2 = (p_1^2+m_1^2)^{\half}+(p_2^2+m_2^2)^{\half} + \kappa |x_1-x_2| \, .
\ee
The coordinates are canonical, i.e. $\{ x_i, p_j \} =   \delta_{ij} $ and all the other brackets vanish.
Later on we will set $\kappa = {   L^2\ov \pi \alpha'} = 2g$ to make contact with the units we used in section \ref{sec:two}. Quantizing   in these coordinates is problematic, because one cannot restrict the relativistic system to the truncated two-particle Fock-space  while preserving Lorentz invariance \cite{Bardeen:1976yt, artru1984, Lenz:1995tj}.

Let us try to evade the anomaly by switching to a different coordinate system. Following \cite{Bars:1975dd, BARS1976413, artru1984, Lenz:1995tj}, we separate center-of-mass and relative motion by
considering the canonical transformation,
\be
  \label{eq:cantran1}
  x_{1,2} = x_0 \pm {x' \ov 2} \, , \qquad  p_{1,2} =  {p_0\ov 2} \pm p' \, ,
\ee
and go to the infinite momentum frame by taking the  $p_0 \to \infty$ limit.
We can write the square of the invariant mass as
\be
  \label{eq:nonsch}
  \Deltax^2 = H_2^2 - p_0^2 \approx 2p_0(H_2-p_0) = {m_1^2 \ov z} + {m_2^2 \ov 1-z} + 2 \kappa|s| \, ,
\ee
where we defined the momentum fraction variable $z$ and the conjugate (signed) action variable $s$,
\be
  \label{eq:momfract}
  z := {p_1 \ov p_1 + p_2} \, , \qquad s := x'(p_1 + p_2) \, ,
\ee
which satisfy $\{ s,z \} =   1 $. Let us now promote \eqref{eq:nonsch} to a \schr equation,
\be
  \label{eq:sch}
  \Deltax^2 \varphi(z) = \le[  {m_1^2 \ov z} + {m_2^2 \ov 1-z} + 2 \kappa|s| \ri] \varphi(z) \, .
\ee
The action of $|s|$ can be found by Fourier transforming the wavefunction, multiplying it by $|s|$, and then Fourier transforming back into $z$-space,
\be
  \nonumber
  |s| \varphi(z)
   = \lim_{\epsilon \to 0} {1\ov {\pi}}  \int_{-\infty}^{\infty}  dz'  \,
   \int_{0}^{\infty}  ds \,   s \cos\le[s (z'-z)\ri] e^{-\epsilon s} \varphi(z')
   = -{1\ov \pi} \dashint_{-\infty}^{\infty} dz' {\varphi(z') \ov (z' - z)^2} \, .
\ee
In order to remove negative energy states, we project on non-negative momenta. The conditions $p_{1,2}\ge 0 $ require $z\ge 0$ and $z\leq 1$, respectively. This restricts the integration domain and thus the \schr equation becomes the 't Hooft equation \eqref{eq:sch} with the identification $\mu^2 = {\pi \Deltax^2 \ov 4g}$. 

\subsection{Canonical variables in flat space}

\label{sec:flat}

The spectral curve of the folded string in AdS$_3$ space is given in \eqref{eq:specu}, which we copy here for convenience,
\be
  \nonumber
   e^p + e^{-p} + 2 - { \Deltax^2 \ov u^2 -4g^2}  = 0 \, .
\ee
Here $p$ and $u$ are canonically conjugate variables. The classical configuration space consists of the union of two half-lines $|u| \ge 2g$. At $u=\pm 2g$, the two particles collide. The flat space limit is achieved by sending $g \to \infty$ while keeping the string energy fixed.

\renewcommand{\arraystretch}{1.2}
 \begin{table}[h!]
\begin{center}
\begin{tabular}[t]{|c|c|l|}
\hline
 {\bf coordinates} &  {\bf description} & \ \quad {\bf Hamiltonian} \\   \hline   \hline
 ($\tilde p, \tilde \rho$) &  physical momentum and position (section \ref{sec:two}) &  $H_\textrm{physical}=|\tilde p| + 4 g |\tilde \rho|$  \\ [1ex]
   ($p, q$) & spectral curve coordinates in the $g\to\infty$ limit & $H_\textrm{RS}=4 \cosh\le({p \ov 2}\ri) \sqrt{ g |q|}$ \\ [1ex]
 ($z, s$)  & parton momentum fraction and action variable  &   $H_\textrm{null}= \sqrt{4g|s|}$
 \\ [1ex]
 \hline
\end{tabular}
\caption{Canonical coordinates for the folded string in flat space.
}
\end{center}
\end{table}

Assuming $u$ is real, we can define a new coordinate
\be
  \label{eq:defqu}
  q := \begin{cases}
  u-2g \qquad & \textrm{for} \ u\geq +2g \, , \\
  u+2g \qquad & \textrm{for} \ u\leq -2g \, .
\end{cases}
\ee
This effectively cuts out the $u\in (-2g, 2g)$ region, which is not part of the classical configuration space.
Let us rescale the energy,
\be
  \nonumber
  \mu^2 := {\pi \Deltax^2 \ov 4g} \, .
\ee
Taking the $g\to \infty$ limit while keeping $\mu$ fixed gives a non-analytic equation,   the (real) spectral curve
\setlength\fboxsep{0.15cm}
\be
  \label{eq:cutspec}
 \boxed{ \quad   e^p + e^{-p}  + 2 - {\mu^2 \ov \pi |q|}  = 0 \, . \quad }
\ee

\begin{figure}[h]
\begin{center}
\includegraphics[width=10cm]{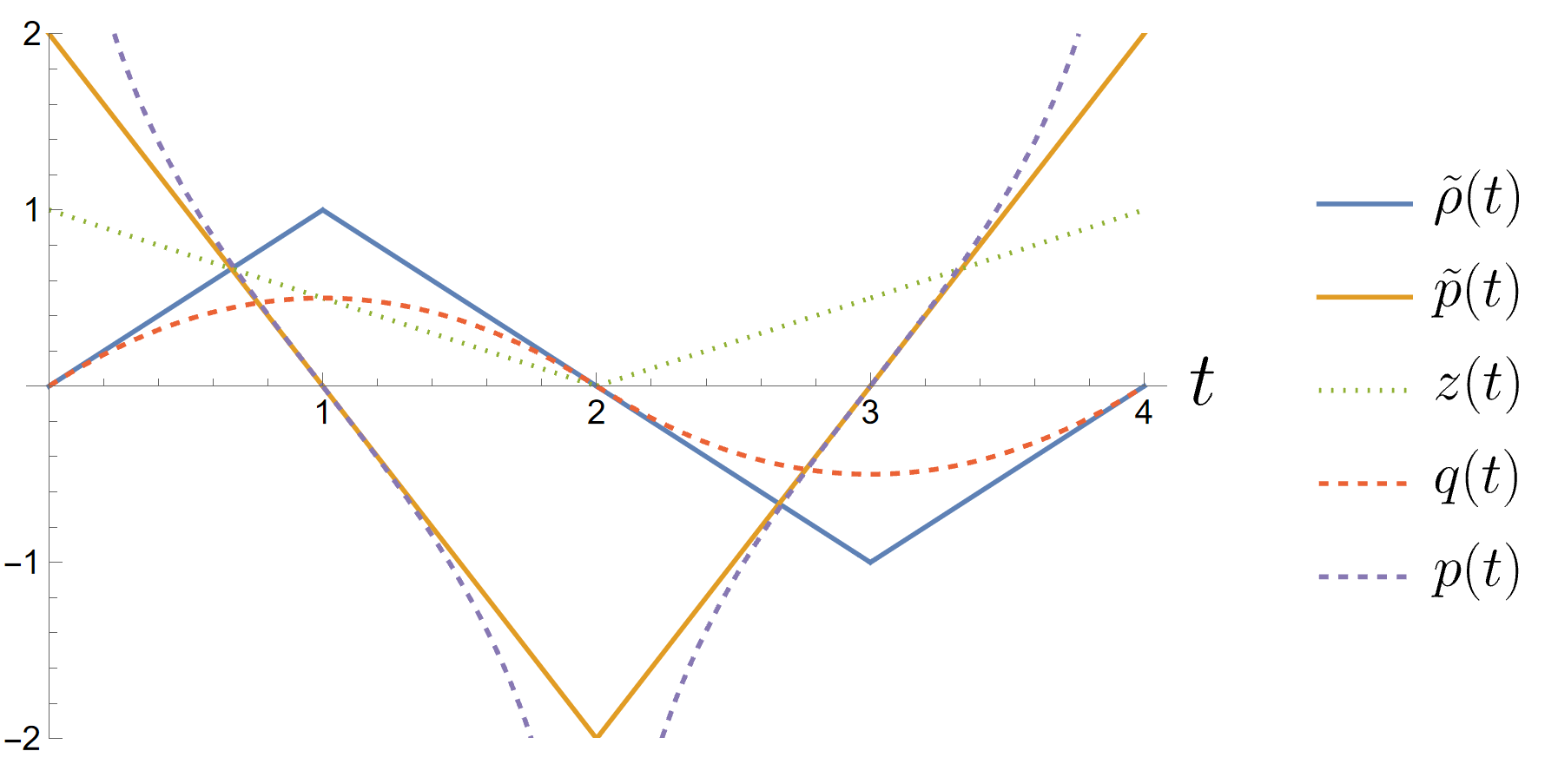}
\caption{\label{fig:flatmotionplot}
Time evolution of dynamical variables of the folded string in flat space ($g=\half$).  Here $\tilde\rho$ is the physical position of one of the particles at the end of the string (see Figure~\ref{fig:yoyo}), $\tilde p$ is the conjugate momentum, $z$ is the momentum fraction variable from eqn. \eqref{eq:momfract}, and $p$ and $q$ denote (non-analytic) spectral curve coordinates; see eqn. \eqref{eq:defqu}.
}
\end{center}
\end{figure}

In the following, we work out the canonical transformations between flat space variables. The discussion is parallel to section \ref{sec:poisson}, where we introduced the relevant tiling variables in the case of AdS target space. Although some of the maps are going to be straightforward limits of their AdS counterparts, they can also be obtained by simple trial and error, because as   Figure \ref{fig:flatmotionplot} shows, the time evolution of the variables is rather simple.
A summary of the relevant canonical variables that we are using is presented in Table~2. Our goal is to determine a direct map between the variables $(z,s)$, which we used in lightcone  quantization, and $(p,q)$, the spectral curve coordinates. This will allow us to convert the 't Hooft equation, which is written in the $z$-basis, into a difference equation in the $q$-basis.

The relationship between $(\tilde p, \tilde \rho)$ and ($p, q$) can be determined from \eqref{eq:cantran} by taking the $g\to \infty$ limit and examining the four possibilities (depending on the signs of $p$ and $q$) on a case-by-case basis. We get
\be
  \label{eq:finrelat}
  p =  (\textrm{sgn} \, \tilde p )  \log\le( 1+  {|\tilde p| \ov 2 g |\tilde \rho|} \ri) \, , \qquad
  q = \half \tilde \rho \le(|\tilde p| + 2 g |\tilde \rho|\ri) \,  ,
\ee
which is a canonical transformation if $\tilde \rho \ne 0$ and $\tilde p \ne 0$.
The Hamiltonian \eqref{eq:hamipu} simplifies to
\be
  \nonumber
    H_\textrm{RS}(p,q)  =  4 \cosh\le({p \ov 2}\ri) \sqrt{ g |q|} \, .
\ee

The solution $q(t)$ consists of parabolas glued together as shown in Figure \ref{fig:flatmotionplot} (red dashed curve).
Setting $m_1 = m_2 = 0$ in eqn. \eqref{eq:nonsch} allows us to express $s$ in terms of $(\tilde p, \tilde \rho)$
\be
  \label{eq:yexpression}
    s =  {H^2 \ov 4g} \textrm{sgn} \, \tilde \rho = {  \textrm{sgn} \, \tilde \rho \ov 4g}\le(|\tilde p| + 4 g |\tilde \rho|\ri)^2 \, .
\ee
The expression includes an extra factor of $\textrm{sgn} \, \tilde \rho$. This  is necessary, otherwise the map  $(\tilde p, \tilde \rho) \mapsto (z, y)$  would be two-to-one.

Note that $s$ is frozen due to the infinite boost and it changes discontinuously whenever $\tilde \rho$ changes sign, i.e. when the quarks collide. Its time evolution is given by the Poisson bracket
\be
  \nonumber
  \dot s = \{s, H\} = {\tilde p^2 \ov 4g} {d\ov d\tilde\rho}  \textrm{sgn} \, \tilde \rho
  = |s| {d\ov d\tilde\rho}  \textrm{sgn} \, \tilde \rho \, ,
\ee

\noindent
which is consistent with the jump. In contrast, $z$ changes continuously between 0 and 1, and we can determine its expression by demanding  $\{\tilde \rho, \tilde{p} \} =  1 $ and    $\{ s,z \} =   1 $. We get
\be
  \label{eq:zexpression}
    z = {  {\vartheta}(\tilde p)\tilde p + 2g |\tilde\rho| \ov  | \tilde p| + 4g |\tilde\rho|}\,  ,
\ee
where $\vartheta(x)$ is the Heaviside step function. This map is a special case of the map  obtained in \cite{Kalashnikova:1997cs}. The Hamiltonian $H = \Deltax$ is independent of $z$,
\be
  \label{eq:flathamis}
  H_\textrm{null}(z,y) = \sqrt{4g|s|} \, .
\ee

Note that the constant $g$ can be eliminated from the maps by rescaling $\tilde\rho \to \tilde\rho/\sqrt{g}$ and $\tilde p \to \tilde p \sqrt{g}$.
Inverting the formulas \eqref{eq:yexpression} and \eqref{eq:zexpression} gives
\be
  \label{eq:inverted}
  \tilde p = 2 (2z-1) \sqrt{g|s|} \, , \qquad
  \tilde\rho = {1\ov \sqrt{ g}}  \le[  \vartheta(2z-1)- z \, \textrm{sgn}(2z-1)  \ri] \sqrt{|s| }\, \textrm{sgn} \, s\, .
\ee
This result finally allows us to relate the coordinate pairs $(z,s) \mapsto (\tilde p, \tilde \rho) \mapsto  (p,q) $. Using \eqref{eq:finrelat} and  \eqref{eq:inverted} we get the simple relationship
\setlength\fboxsep{0.15cm}
\be
  \label{eq:pqzy}
 \boxed{ \quad p = \log{z \ov 1-z} \, , \qquad q =   s z(1-z) \, . \quad }
\ee
The map is a canonical transformation between $0 < z < 1$ and singular at the boundaries of the interval where $p$ diverges.
Note that the first expression, along with the definition $z \equiv {p_1 \ov p_1 + p_2} $,
implies that  $p$ can be interpreted as the rapidity difference between the quark and antiquark, i.e. the quark momenta can be written as $p_{1,2}= \bar p e^{\pm {p\ov 2}}$.

\subsection{Quantum spectral curve in flat space}

In this section, we will recall how the 't Hooft equation can be converted into a finite difference equation \cite{Fateev:2009jf}.
Since the folds of the yoyoing string in Figure \ref{fig:yoyo} move with the speed of light, we will henceforth  set the renormalized quark masses in \eqref{eq:thooft} to zero. Hence, our starting point is the equation,
\be
  \label{eq:thooft2}
   \mu^2 \varphi(z) =  - \dashint_{0}^1 dz'  {\varphi(z') \ov (z' - z)^2} \, ,
\ee
which we would like to relate  to the classical spectral curve of the folded string.

\subsubsection*{Boundary layer equation}

In order to calculate the wavefunctions by a matched asymptotic expansion, the authors of \cite{Brower:1978wm} considered the boundary layer near $z=0$. By means of a Fourier transformation, they derived a difference equation from \eqref{eq:thooft2}, which we will now review as a warm-up exercise. Let us define the boundary layer variable $\xi = z \mu^2 $. This eliminates $\mu$ from the equation,
\be
  \nonumber
  \varphi(\xi) =   - \dashint_{0}^{\infty} d\eta  {\varphi(\eta) \ov (\eta - \xi)^2} \, .
\ee
After changing to rapidity coordinates $\xi = e^\theta $, the 't Hooft equation becomes
\be
 \nonumber
     {\varphi(\theta)}  = -{e^{-\theta}\ov 4}  \, \dashint_{-\infty}^{\infty} d\theta'  {\varphi(\theta') \ov \sinh^2\le({\theta-\theta' \ov 2}\ri)} \, .
\ee
After multiplying both sides by $e^{\theta}$ and performing a Fourier transformation\footnote{Note that a Fourier transformation in rapidity coordinates is equivalent to a Mellin transformation.}
\be
  \nonumber
  \chi(\nu) = {1\ov \sqrt{2\pi}} \int_{-\infty}^{\infty}  d\theta \, e^{i\nu \theta} {\varphi(\theta)}  \, ,
\ee
we get
\be
  \label{eq:brower}
   \chi(\nu-i) =\pi \nu \coth\le({\pi \nu } \ri) \chi(\nu)  \, .
\ee
The equation can be solved in terms of the Barnes G-function \cite{Fateev:2009jf}.

\subsubsection*{Quantum spectral curve}

Following \cite{Fateev:2009jf}, we can also convert \eqref{eq:thooft2} into a finite difference equation without restricting our attention to a boundary layer. Let us first switch to spectral curve momentum coordinates $z \to p$ via the transformation that we have found in \eqref{eq:pqzy}. This change of variables has also been used in \cite{Narayanan:2005gh}.
The 't Hooft equation becomes
\be
  \nonumber
   \mu^2{\varphi(p)}  = - \cosh^2 \le({p\ov 2}\ri) \, \dashint_{-\infty}^{\infty} dp'  {\varphi(p') \ov \sinh^2\le({p-p'\ov 2}\ri)} \, .
\ee
After dividing by $4\cosh^2 \le({p\ov 2}\ri)$ and Fourier transforming $ \varphi(p) \to \psi(\nu)$, we get
\be
   \nonumber
 {\mu^2 \ov  2\pi} \int_{-\infty}^{\infty} d\nu' S(\nu - \nu') \psi(\nu') = \pi \nu \coth\le({\pi \nu  } \ri) \psi(\nu)  \, ,
\ee
where the kernel $S$ is given by
\be
  \nonumber
  S(\nu) = {\pi \nu \ov   \sinh\le({\pi \nu } \ri) } \, .
\ee
Let us define the $Q$-function \cite{Fateev:2009jf}
\be
  \label{eq:defq}
  Q(\nu) := \nu  \cosh\le({\pi \nu  } \ri) \psi(\nu)
   \, .
\ee
Then for the strip $-1 < \Im \nu < 1$ we have
\be
  \nonumber
  Q(\nu) =   {  \mu^2 \ov 2\pi}  \sinh\le({\pi \nu  } \ri)  \dashint_{-\infty}^{\infty} d\nu' \, {\nu - \nu' \ov \nu'} \cdot {1\ov \cosh {\pi \nu' }   }\cdot {1\ov \sinh {\pi (\nu-\nu') }  }Q(\nu') \, .
\ee
If we shift the argument, then according to the  Sokhotski–Plemelj theorem (appropriately modified for the strip; see e.g.  \cite{Antipov} for a similar calculation), we get an extra term on the boundary
\be
  \nonumber
  Q(\nu \pm  i) =  -{ \mu^2 \ov 2\pi} { \tanh({\pi \nu  }  )  \ov  \nu   }  Q(\nu)+  {  \mu^2 \ov 2\pi}  \sinh\le({\pi \nu  } \ri) \dashint_{-\infty}^{\infty} d\nu' \, {\nu \pm i - \nu' \ov \nu'} \cdot {1\ov \cosh {\pi \nu'  }   }\cdot {1\ov \sinh {\pi (\nu-\nu')  }  }Q(\nu') \, .
\ee
Using these expressions one can show that $Q(\nu)$ satisfies the following difference equation,\footnote{Note that the boundary layer  difference equation \eqref{eq:brower} can be recovered from \eqref{eq:findiff} by plugging in $Q(\nu)~=~\nu (-\mu^2 )^{-{i \nu  }} \chi(\nu)$ and then taking the large $\mu^2$ limit.}
\setlength\fboxsep{0.15cm}
\be
  \label{eq:findiff}
 \boxed{ \quad   Q(\nu+i)+Q(\nu-i) -2 Q(\nu) = -{\mu^2 } { \tanh({\pi \nu}  )  \ov  \pi\nu   } Q(\nu)  \, . \quad }
\ee
This equation\footnote{Note the factor of two difference between the definition of $\nu$ in \cite{Fateev:2009jf} and our notation.} was first derived in \cite{Fateev:2009jf}. The authors have developed an analytic method to  determine the meson mass spectrum using a systematic expansion. For further details, we refer the reader to \cite{Fateev:2009jf}.

Our claim is that it can be interpreted as the quantized spectral curve of the folded string in Figure \ref{fig:yoyo}.
The classical spectral curve of the string is given in \eqref{eq:cutspec}, which we copy here for convenience,
\be
  \label{eq:copycon}
 e^p + e^{-p}  + 2 - {\mu^2 \ov \pi |q|}  = 0 \, .
\ee
After identifying $\nu = q$, the relationship of \eqref{eq:findiff} and \eqref{eq:copycon} are clear.
Let us assume that the state is described by a wavefunction $\psi(q)$, which vanishes if we act on it by an (appropriately defined) operator version of \eqref{eq:copycon}.
Taking the representation $\hat p= -i \p_{q}$ and $\hat q = q\cdot$, the  exponential  of the momentum becomes a shift operator
\be
  \nonumber
  e^{\pm p} \psi(q) = \psi(q \pm i) \, .
\ee
In the classical (large $\nu$) limit, we have  $\tanh({\pi \nu }) \to \textrm{sgn} \, \nu$, and \eqref{eq:findiff} reduces to \eqref{eq:copycon}  modulo a minus sign multiplying $e^p + e^{-p}$. This is due to the definition of $Q$ in  \eqref{eq:defq}, which contained the factor $\cosh\le({\pi \nu } \ri)$, an antiperiodic function under the shift $\nu \to \nu + i$. We emphasize that at this point it is unclear how to directly quantize  \eqref{eq:copycon} and why we must replace $|q| \mapsto q \coth \pi q$. We leave this for future work.

\subsection{Quantum spectral curve in AdS }

The results of the previous section   can be extended to AdS by noticing that in the $g\to \infty$ limit we can relate  flat space  coordinates to tiling variables in AdS,
\be
  \nonumber
  (A,B) = {x' \ov 4g} (p_1, p_2) \, ,
\ee
where $x'$ is the relative position and $p_1, p_2$ are the particle momenta from \eqref{eq:cantran1}.
In fact, with the above identification the Poisson bracket \eqref{eq:linpoi} follows from  $\{ x_i, p_j \} =   \delta_{ij} $ even away from $g=\infty$.
Furthermore, we have
\be
  \nonumber
  (Z,S)  = (z,s)\, .
\ee
Moreover, the coordinate transformation \eqref{eq:pqzy}, which is valid in flat space, is consistent with its AdS counterpart given in \eqref{eq:pqzy0} and \eqref{eq:pqzy00} for the two maps.
Finally, the Hamiltonian \eqref{eq:flathamis} is the flat space limit of \eqref{eq:abshami}.
Having identified the tiling variables and the Hamiltonian as a one-parameter deformation of the flat space system, we can now promote them to operators.

\subsubsection*{The case of $g=0$}

This case is simple, because the first and second maps coincide and we do not need to take absolute values.
First, setting $g=0$ in the classical spectral curve \eqref{eq:specu} gives
\be
  \label{eq:freeclassspec}
   e^p + e^{-p} + 2 - { \Deltax^2 \ov u^2 }  = 0 \, .
\ee
On the other hand, by taking the square of \eqref{eq:abshami}  we obtain
\be
  \label{eq:free}
  \Deltax^2 \varphi(z) =  \hat S \hat Z(1- \hat Z) \hat S \varphi(z)  \, ,
\ee
where we have symmetrized the operator on the RHS, which acts on the wavefunction $\varphi(z)$ where $z \in [0,1]$. Taking the representation $ \hat Z = z\cdot$ and $\hat S = -i \p_z$, we obtain
\be
 \label{eq:spinequ}
 z(1-z)\varphi''(z) + (1-2z) \varphi'(z) + \Deltax^2 \varphi(z) =  0 \, .
\ee
The term on the RHS of \eqref{eq:free} has  been proposed as  an effective  confining potential to capture the nonperturbative dynamics in hadron physics~\cite{Li:2015zda}.
Eqn. \eqref{eq:spinequ} has been analyzed in \cite{Faddeev:1994zg} in the context of the XXX Heisenberg spin chain for zero spin and two sites.  The corresponding Baxter equation in  \cite{Faddeev:1994zg} is nothing but the quantized classical curve \eqref{eq:freeclassspec}.

Note that in terms of the momentum variable $p=\log {z\ov 1-z}$  the equation is cast into a scattering equation with the  P\" oschl--Teller potential,
\be
  \nonumber
  \varphi''(p) + {\Deltax^2 \ov 4 \cosh^2\le({p\ov 2}\ri)}\varphi(p) = 0 \, ,
\ee
and it is well-known that the potential is reflectionless precisely if
\be
  \label{eq:spinspec}
  \Deltax^2 = n(n+1) \, , \qquad n = 0,1,2,\ldots \, .
\ee
This formula also gives the spectrum.
Notice that the calculation does not require anything extraordinary, since in the  $g= 0$ case, the physical domain of $u$ is the entire real line; therefore no issues arise regarding the definition of the $p$-basis.

Note that we could have symmetrized the operator $Z(1-Z)S^2$  in the opposite way.
The relationship between the two ordering is given by
\be
  \nonumber
    [\hat Z(1- \hat Z) ]^\half \hat S^2  [\hat Z(1- \hat Z) ]^\half  \varphi(z)- \hat S \hat Z(1- \hat Z) \hat S \varphi(z)
    = {1\ov 4} \le[ {1 \ov z} + {1 \ov 1-z} \ri] \varphi(z) \, ,
\ee
which is proportional to the mass term in the 't Hooft equation \eqref{eq:thooft}. If we multiply the RHS by $m^2$ and add it to the RHS of \eqref{eq:free}, then the equation can still be solved analytically in terms of Jacobi polynomials  \cite{Li:2015zda}. The corresponding eigenvalues are
\be
  \nonumber
  \Deltax^2 = (n+|m|)(n+|m|+1) \, , \qquad n = 0,1,2,\ldots \, .
\ee

\subsubsection*{Finite $g$  spectral curve}

At finite $g$ values, the 't Hooft equation acquires another term (again we change $Z \to z$),
\setlength\fboxsep{0.25cm}
\be
   \label{eq:adsequation}
  \boxed{ \quad  \mu^2 \varphi(z) =  - \dashint_{0}^1 dz'  {\varphi(z') \ov (z' - z)^2} -  {\pi \ov 4g} \le[ z(1-z)\varphi''(z) + (1-2z) \varphi'(z) \ri] \, .  \ \ }
\ee
After switching to the $p$  coordinate we have
\be
   \mu^2{\varphi(p)}  = - \cosh^2 \le({p\ov 2}\ri) \, \dashint_{-\infty}^{\infty} dp'  {\varphi(p') \ov \sinh^2\le({p-p'\ov 2}\ri)} - {\pi \ov g }\cosh^2\le({p\ov 2}\ri) \varphi''(p)  \, .
\ee
Similarly to the previous section we now perform a Fourier transformation and switch to the Q-function defined by,
\be
  \nonumber
  Q(\nu) := \le[ \nu  \cosh\le({\pi \nu  } \ri) + {\nu^2 \ov 4g}  \sinh\le({\pi \nu  } \ri)  \ri]\psi(\nu)
   \, .
\ee
By naively repeating the same steps as before we get the difference equation,
\setlength\fboxsep{0.15cm}
\be
  \label{eq:findiffg}
 \boxed{ \quad   Q(\nu+i)+Q(\nu-i) -2 Q(\nu) = -  {\mu^2   \ov {\pi \nu^2 \ov 4g}+ \pi\nu   \coth({\pi \nu) }  } Q(\nu)  \, . \ \ }
\ee
This is to be compared with the classical spectral curve \eqref{eq:specu} in the $q$ coordinate defined by \eqref{eq:defqu}. In this case we do not take the large-$g$ limit so that  we obtain the following non-analytic spectral curve,
\setlength\fboxsep{0.15cm}
\be
 \label{eq:gennonanal}
     e^p + e^{-p}  + 2 - {\mu^2 \ov {\pi q^2 \ov 4g}+ \pi |q|}  = 0 \, .
\ee
Indeed, by comparing \eqref{eq:findiffg} and \eqref{eq:gennonanal} we see that the difference equation can be interpreted as the quantum spectral curve.

It should be mentioned that eqns. \eqref{eq:findiff} and \eqref{eq:findiffg} are different from the quantum spectral curves that appear in the literature in a crucial aspect, namely that their classical limit is not analytic. The reason for this is that the classical phase space of the folded string in the $(z,s)$ coordinates is mapped to the spectral curve coordinates $(p,u)$ via the ``mixed map'' of section \ref{sec:poisson}, which contained absolute values and was only defined for $u \in \RR$.

\clearpage

\section{Numerical spectrum}

\label{sec:numerics}

In this section we will numerically solve the modified 't Hooft equation \eqref{eq:adsequation}, which governs the wavefunction of the internal degree of freedom of the four-segmented string in AdS$_3$  (or that of the folded string in AdS$_2$).  The equation is an eigenvalue equation for $\mu$ and it contains a single parameter $g$, which sets the size of AdS. The spectrum is already known in two different limits. As $g\to \infty$, we recover the 't Hooft equation, which has been studied extensively starting with \cite{tHooft:1974pnl, HANSON1977477, Brower:1978wm}.
On the other hand, by plugging in $\mu^2 \to {\pi \Deltax^2 \ov 4g}$ and taking $g=0$ one obtains \eqref{eq:spinequ} and  the spectrum should match the result in eqn. \eqref{eq:spinspec}.
Following \cite{Brower:1978wm}, let us define a new coordinate $\theta$ by setting $z~=~\half(1-\cos \theta)$ and expand the wavefunction
\be
  \nonumber
  \varphi(z) = \sum_{k=1}^\infty  a_k \sin k \theta \, .
\ee
By plugging this into the modified 't Hooft equation \eqref{eq:adsequation}, multiplying by ${2 \ov \pi}\sin l \theta$ and integrating w.r.t. $\theta$, we arrive at a matrix eigenvalue problem,
\be
  \nonumber
  \mu^2 a_k = (V+ {\pi \ov 4g }U)_{kl} \, a_l
\ee
where for the elements of $V$ and $U$ we obtain
\be
  \nonumber
  V_{kl} = -{4 \ov \pi} \int_0^\pi d\theta  \sin k \theta \,  \dashint_0^\pi d\theta' { \sin \theta' \sin l \theta'  \ov (\cos \theta -\cos \theta')^2}
  = 4 l \int_0^\pi d\theta {\sin k \theta \sin l \theta \ov \sin \theta }  \, ,
\ee
\be
  \nonumber
 U_{kl} ={2 l \ov \pi} \int_0^\pi  d\theta \le[l\sin l \theta - \cos l \theta \cot \theta \ri] \sin k \theta \, .
\ee
The integrals can be evaluated explicitly; see e.g. Appendix A in \cite{Lebed:2000gm}.
For the first column of $V$ we get,
\be
  \nonumber
  V_{k,1} = \begin{cases}
   {8 \ov k} \qquad & \textrm{for} \   k = \textrm{odd}\, , \\
 0 \qquad & \textrm{for} \  k = \textrm{even}  \, .
\end{cases}
\ee
For $l>1$ we have,
\be
  \nonumber
  V_{kl} = \begin{cases}
  0 \qquad & \textrm{for} \ k+l = \textrm{odd} \, , \\
    {8l \ov k+l-1}+ {l\ov l-1} V_{k-1,l-1} \qquad & \textrm{for} \  k+l = \textrm{even} \, ,
\end{cases}
\ee
where $V_{0,l} = V_{k,0} = 0$.
In order to compute $U_{kl}$ notice that whenever the integral does not trivially vanish due to the symmetries of the integrand, then $ \cos l \theta \cot \theta  \sin k \theta $ can be written as a $\sum_n c_n \cos(2n\theta)$ and only the constant term $c_0 = 0, \half $ or 1 contributes to the integral.
We obtain,
\be
  \nonumber
  U_{kl} = \begin{cases}
  0 \qquad & \textrm{for} \ k<l \ \textrm{or} \ k+l = \textrm{odd} \, , \\
  {  k(k-1)  }  \qquad & \textrm{for} \   k=l  \, , \\
  -2 l \qquad &  \textrm{otherwise} \, .
\end{cases}
\ee

\renewcommand{\arraystretch}{1.0}
 \begin{table}[h!]
\begin{center}
\begin{tabular}[t]{|c|c|}
\hline
\ $n$ \quad &  {\bf $\mu_n^2 / \pi^2$}  \\   \hline   \hline
 1 &  0.737061746  \\
 2 &  1.753731337  \\
 3 &  2.748160912 \\
 4 &  3.751057582  \\
 5 &  4.749295381  \\
 \hline
\end{tabular} \qquad\qquad
\begin{tabular}[t]{|c|c|}
\hline
\ $n$ \quad &  {\bf $\mu_n^2 / \pi^2$}  \\   \hline   \hline
 6 &  5.75049262  \\
 7 &  6.74962942  \\
 8 &  7.75028440 \\
 9 &  8.74977158  \\
 10 & 9.75018514  \\
 \hline
\end{tabular}
\caption{The first ten eigenvalues at $g=\infty$ computed using $500\times 500$ matrices. The computation time on a laptop was about one minute. The results are in agreement with the literature; see for instance \cite{Fateev:2009jf}.
}
\end{center}
\end{table}

\begin{figure}[h]
\begin{center}
\includegraphics[width=8.5cm]{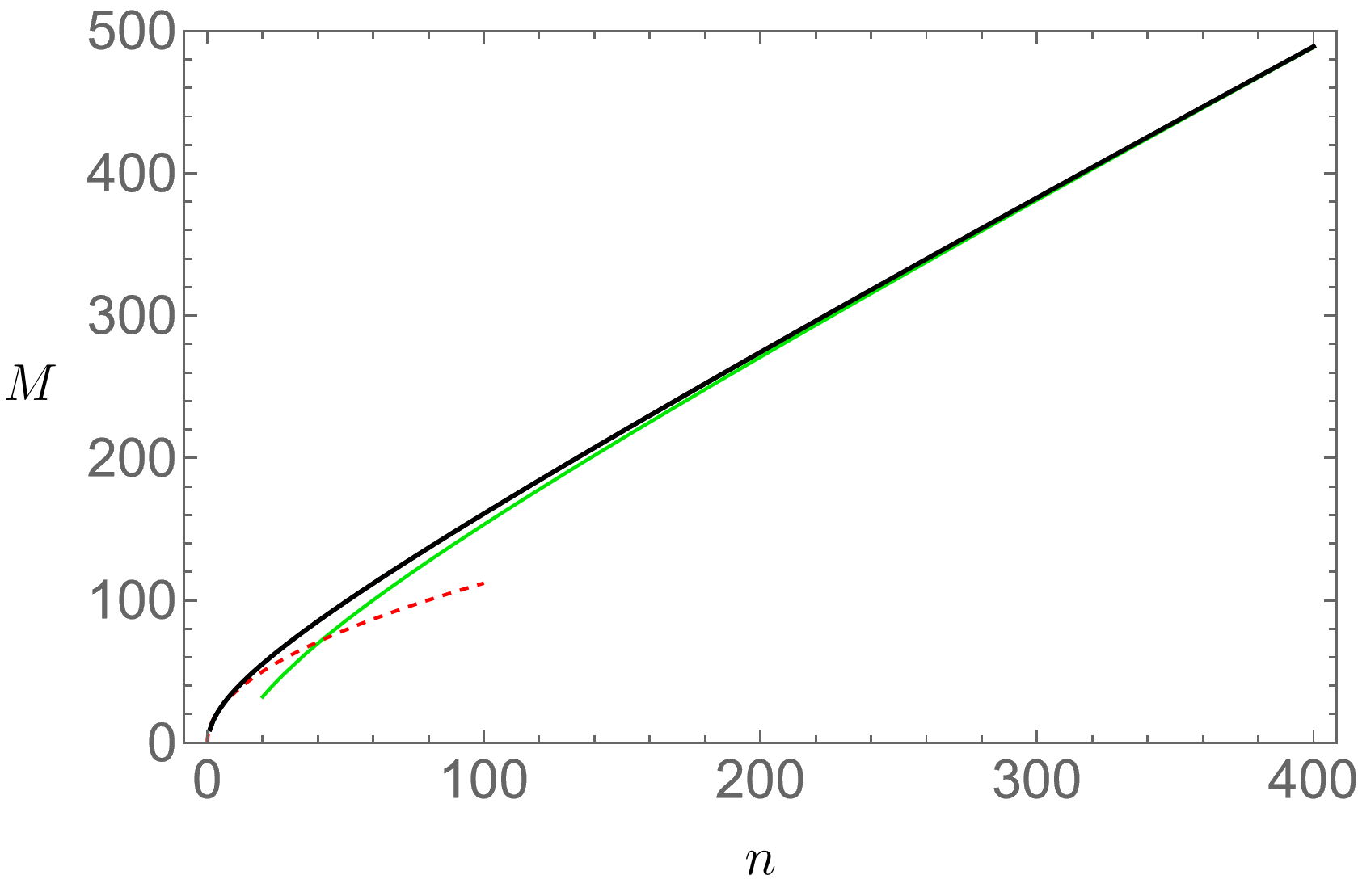}
\caption{\label{fig:gten}
The first 400 eigenvalues at $g=10$. Green line indicates the semiclassical result $\Deltax   = n+  {8g \over \pi } \log n + \textrm{const.}$ from eqn. \eqref{eq:logar}. Red dashed line shows the Regge trajectory  \eqref{eq:regge}.
}
\end{center}
\end{figure}

Note that the $U$ and $V$ matrices are not Hermitean, because we used a basis which is orthogonal with a different inner product. Nevertheless, the sum $V+ {\pi \ov 4g }U$ has real and positive eigenvalues when $g\geq 0$.
In our basis, $U$ is an upper triangular matrix, whose eigenvalues equal the values in the diagonal, thereby confirming the $g=0$ result \eqref{eq:spinspec}.

For practical numerical calculations we will truncate the matrices. With a size of $500\times 500$, one is able to obtain several significant digits, and the calculations typically only take a few minutes on a laptop. The first ten (squared) energy eigenvalues at $g = \infty$ are given in Table~3. This result is well known in the literature.

Figure \ref{fig:gten} shows the eigenvalues at $g=10$ where the modifying term in \eqref{eq:adsequation} (which is proportional to $1/g$)  starts playing a role. With decreasing $g$, the spectrum becomes more and more linear and approaches the $g=0$ result \eqref{eq:spinspec}.

\begin{figure}[h]
\begin{center}
\includegraphics[width=7.5cm]{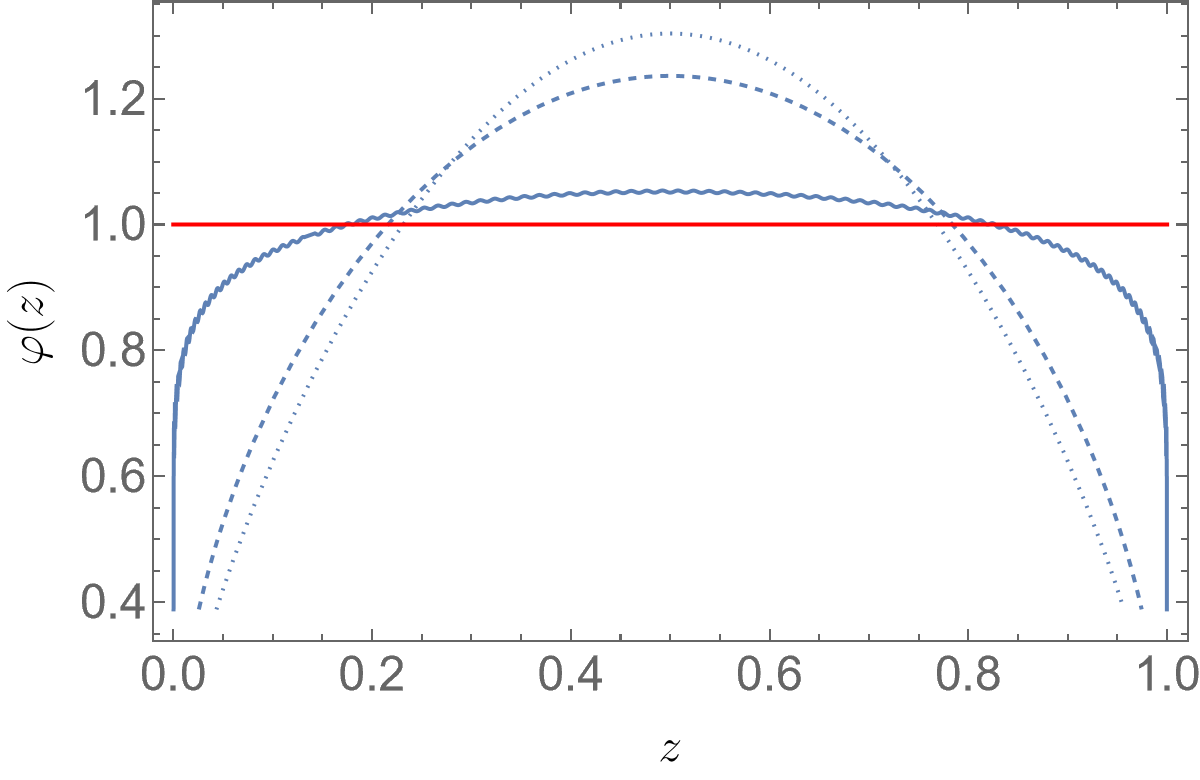}
\includegraphics[width=7.5cm]{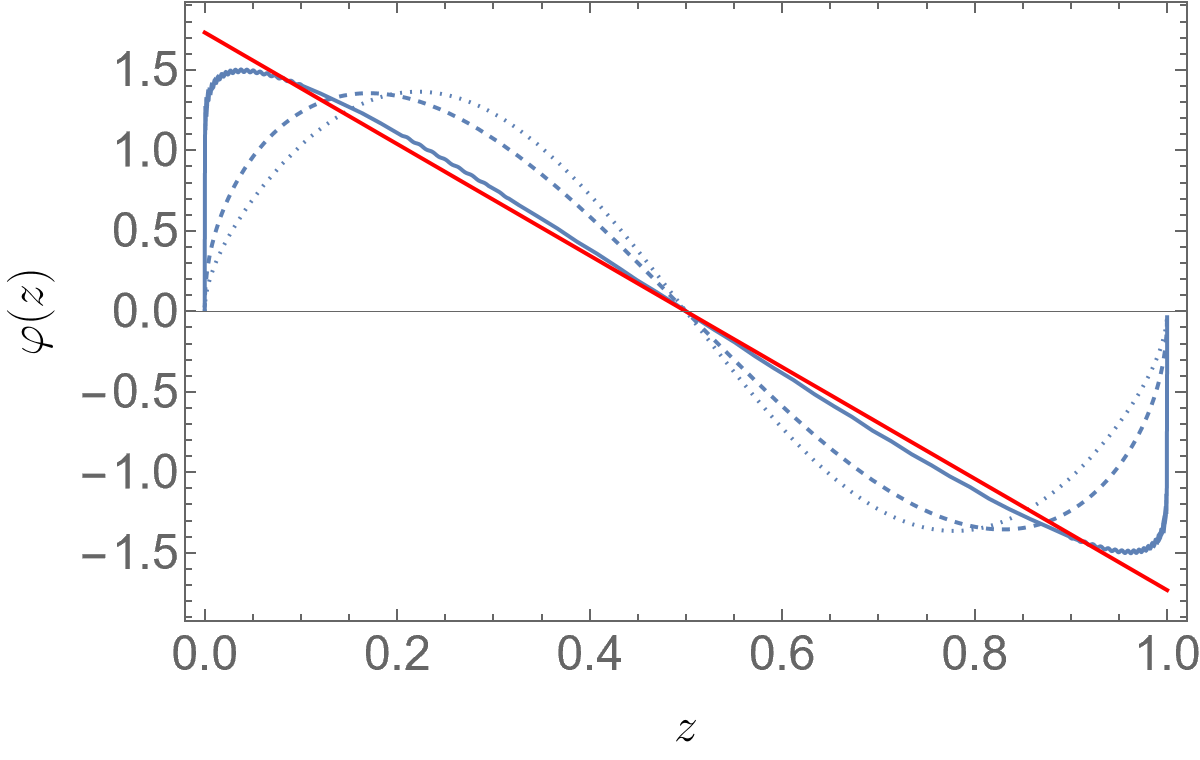}
\includegraphics[width=7.5cm]{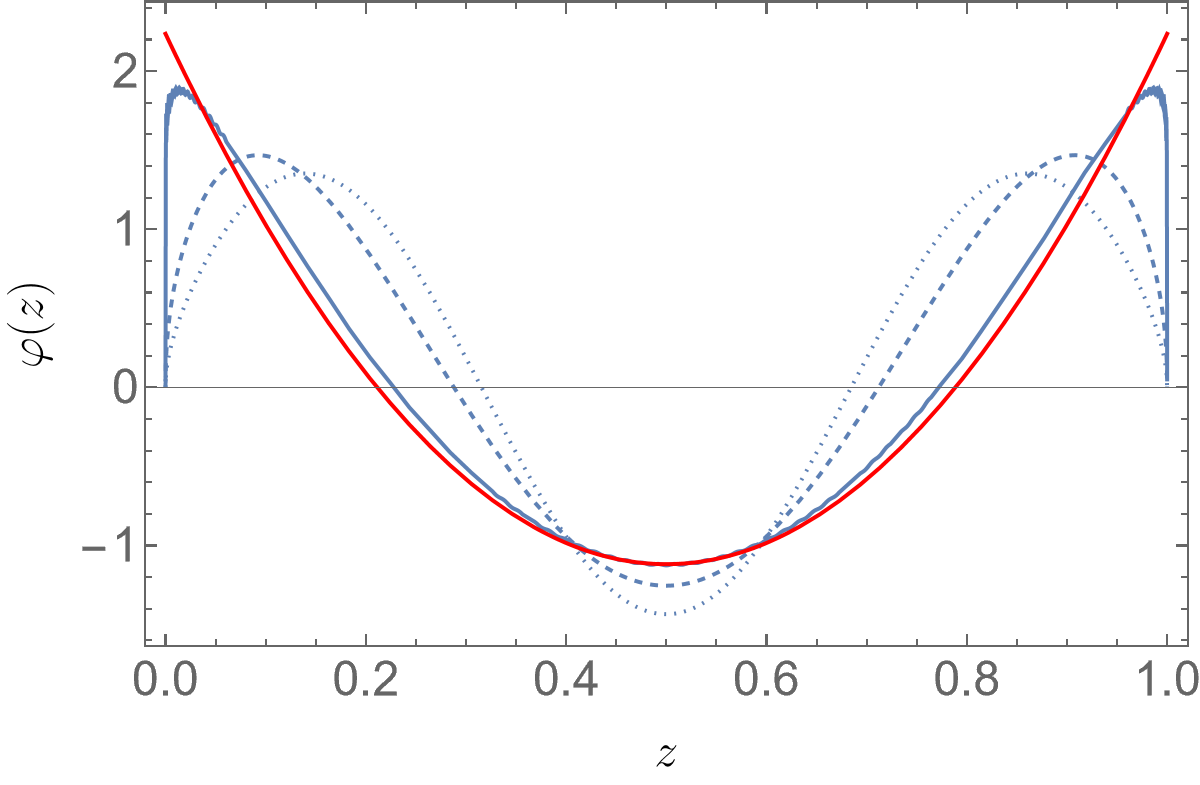}
\includegraphics[width=7.5cm]{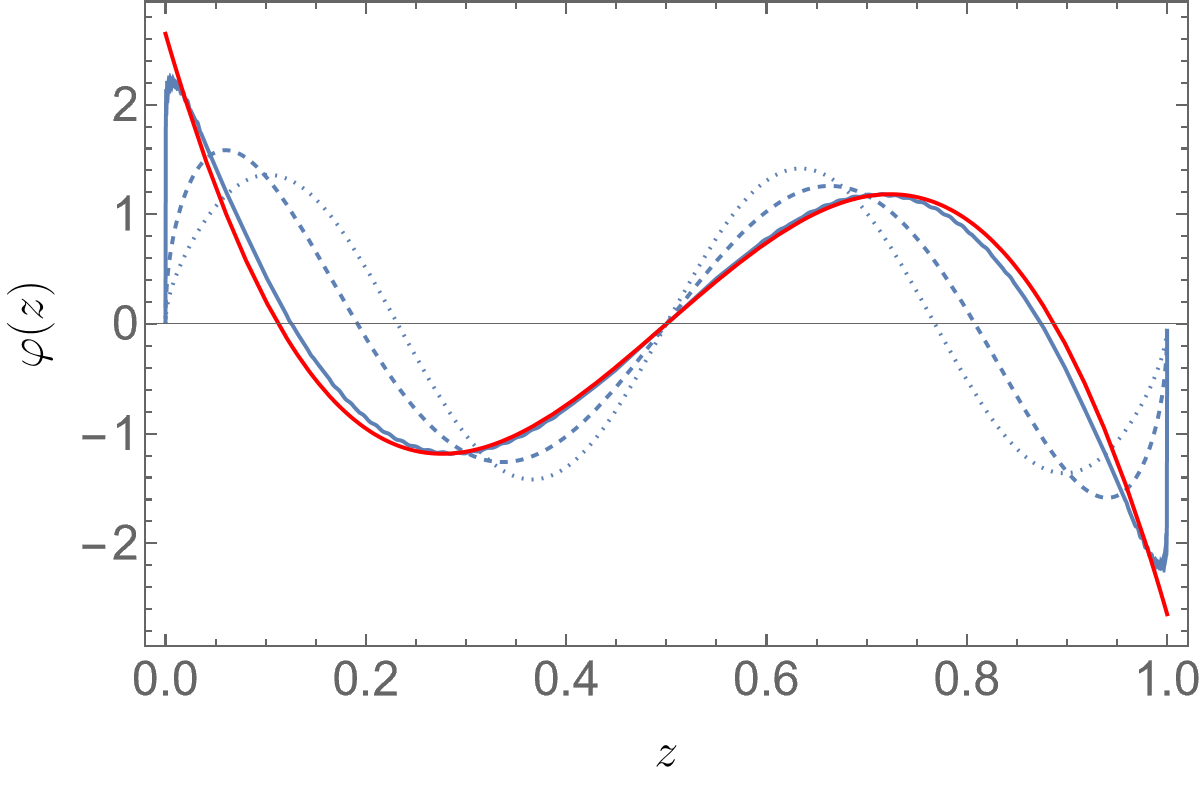}
\caption{\label{fig:wfs}
The first four eigenfunctions $\varphi^{(n)}(z)$ at $g={1\ov 100}$ (solid line), $g={1\ov 2}$ (dashed line), and $g={100}$ (dotted line). Red curves show the analytical result at $g=0$.
}
\end{center}
\end{figure}

Figure \ref{fig:wfs} shows the wavefunctions of the four lowest lying states at various values of~$g$. The analytical form of the wavefunction is only known at $g=0$, in which case it can be expressed in terms of Legendre polynomials. Equation \eqref{eq:adsequation} is symmetric under the exchange $z \mapsto 1-z$ and thus the wavefunctions are either symmetric or antisymmetric. The only qualitative difference between $g=0$ and $g>0$ is the behavior of the wavefunctions near $z=0, 1$. These points correspond to the quark collision events (i.e. $\tilde\rho=0$ in Figure~\ref{fig:yoyo}).

\begin{figure}[h]
\begin{center}
\includegraphics[width=12cm]{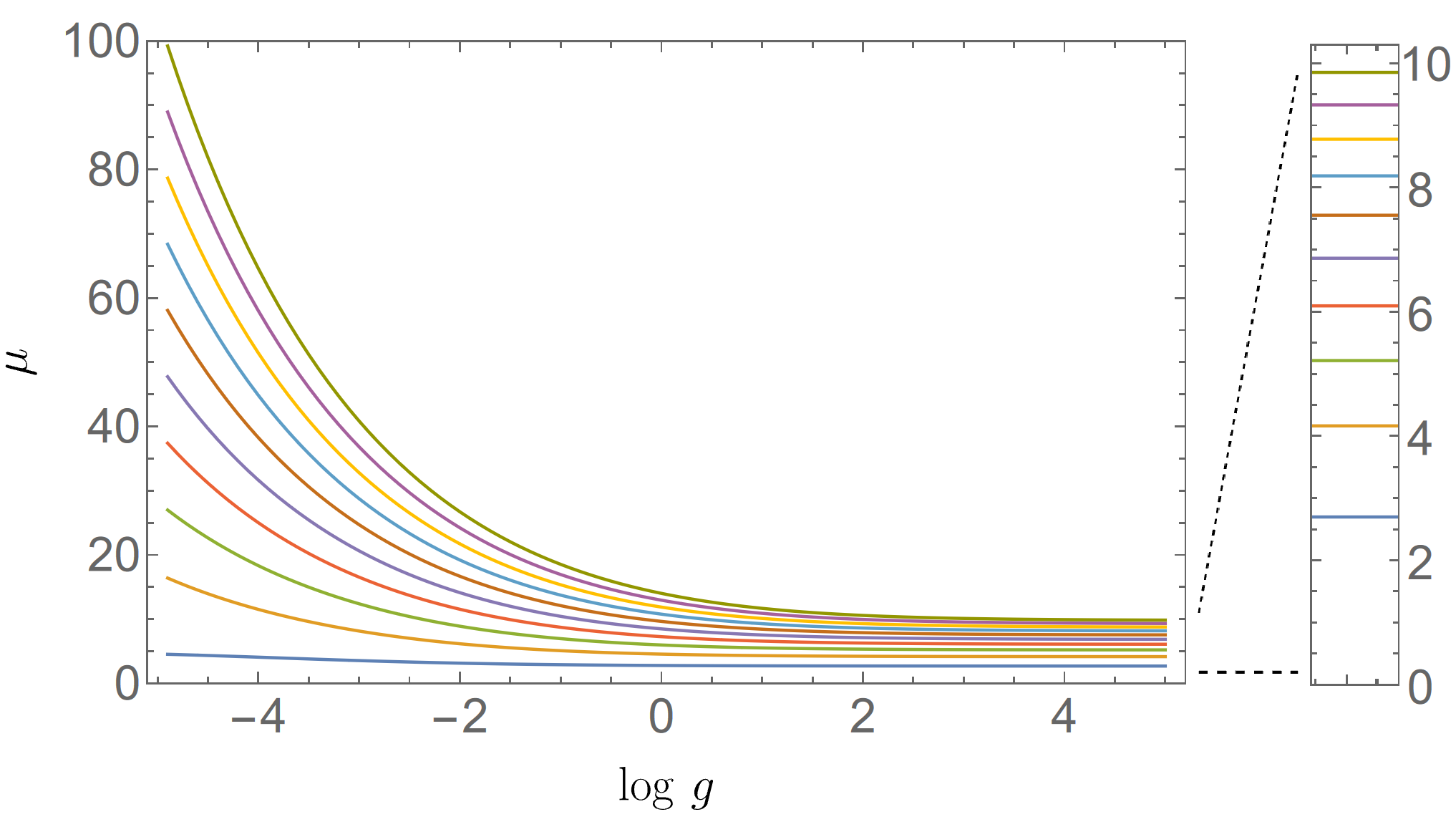}
\caption{\label{fig:spectrum}
The first ten energy eigenvalues of the modified 't Hooft equation plotted as a function of $g={L^2 \ov 2 \pi \alpha'}$ (here $L$ is the AdS radius). The spectrum interpolates between approximately equal spacing at low $g$ and Regge behavior at large $g$.
}
\end{center}
\end{figure}

Finally, Figure~\ref{fig:spectrum} shows the first ten energy eigenvalues as a function of $g$. At $g=\infty$, eqn. \eqref{eq:adsequation} reduces to the 't Hooft equation, whose spectrum lies on a nearly linear Regge trajectory. The plot interpolates between this spectrum and the  $\Deltax_n^2=n(n+1) $ result at $g=0$. Since $g \propto L^2$, from the perspective of the four-segmented string changing $g$ is tantamount to changing the AdS radius. Note that the $g=0$ case is the highly quantum tensionless limit ($\alpha' \to \infty$). Interestingly, the spectrum seems to be compatible with  expectations based on a hypothetical free holographic dual theory with evenly spaced operator dimensions.

\section{Discussion}

In this paper, we attempted to  establish a link between lightcone quantization and quantum spectral curves. In order to clarify their relationship, we considered the simple system of a folded string, which is depicted in Figure \ref{fig:yoyo}.  The one-directional motion of the string  is oscillatory and the endpoints always move with the speed of light.
In flat space, the system is amenable to lightcone quantization, which uses an infinitely boosted frame and a projection on non-negative particle momenta, thereby avoiding a \poincare anomaly in the quantum theory \cite{Lenz:1995tj}. The center-of-mass coordinates decouple and canonical quantization of the fractional momentum and conjugate action variables lead to the 't Hooft equation.

From a certain perspective, the folded string is a degenerate system, because it contains both particles and string segments. Furthermore, having a flat target space is slightly more complicated from a spectral curve point of view. For this reason, we generalized the setup and considered a four-segmented string in AdS$_3$ (see Figure \ref{fig:phases}), which in a squashed limit reduces to the folded string.
We investigated its phase space and computed its classical spectral curve by means of planar bipartite graphs (a.k.a. brane tilings or on-shell diagrams).
As it turns out, the phase space of the internal degree of freedom of the four-segmented string  in AdS$_3$ is identical to that of the folded string in AdS$_2$. We found a canonical pair of ``separated'' variables on the spectral curve, denoted by $(p,u)$.  Physical positions of the particles at the end of the folded string can be obtained from certain level sets, which is familiar from the way one computes soliton positions in 2d relativistic integrable theories from the associated Ruijsenaars-Schneider model.

Using the Baker-Akhiezer function, we managed to map $(p,u)$ onto the natural tiling variables $(Z,S)$ in two different ways, dubbed the {\it first map} and the {\it second map}. These are asymmetric maps, which correspond to the two half periods of the oscillation. The correct map to the variables that appear in the 't Hooft equation turned out to be a certain combination of the two maps, termed the {\it mixed map}.

These maps between phase spaces allowed us to identify the integral transformed 't Hooft equation \eqref{eq:findiff} (first calculated in \cite{Fateev:2009jf}) as the quantum spectral curve of the folded string. Moreover, our findings suggested a natural generalization of the results to four-segmented strings in AdS$_3$. In AdS, the  't Hooft equation is modified by an extra term,
\be
 \label{eq:finalth}
   \mu^2 \varphi(z) =  - \dashint_{0}^1 dz'  {\varphi(z') \ov (z' - z)^2} +  {\pi \ov 4g} (-i\p_z) z(1-z) (-i\p_z) \varphi(z) \, ,
\ee
where $g$ is the square radius of AdS in units of $2\pi\alpha'$. In the context of hadron physics, this term has  previously been proposed as an effective confining potential to capture the non-perturbative dynamics of QCD  \cite{Li:2015zda}. Here it arises as a contribution from the curvature of the AdS target space. It is remarkable that this simple additive term captures all the differences between flat space and AdS. We have solved the equation numerically for various values of $g$ and showed that the spectra interpolate between  $\mu^2={\pi \ov 4g}n(n+1)$ in the $g\to 0$  limit and 't Hooft's nearly linear Regge trajectory as $g \to \infty$. The  $g\to 0$  limit of \eqref{eq:finalth} has been analyzed in \cite{Faddeev:1994zg} in the context of a two-site spin chain with zero spins. It would be interesting to understand the folded string from a spin chain perspective.

A few comments regarding the quantum spectral curve are in order. We have derived the difference equation \eqref{eq:findiff} by a transformation applied to the 't Hooft equation and at this point it is unclear how to obtain it by a direct quantization of the classical curve \eqref{eq:cutspec}.
A peculiar feature of the quantum spectral curve is that its classical counterpart is non-analytic. The non-analyticity arises from the fact that the lightcone coordinates and the spectral curve coordinates are related by the mixed map, which switches between the two canonical maps   depending on the sign of $u$.

We note that it seems possible to quantize the folded string using either the first or the second map instead of the mixed map. The situation is complicated by the fact that the naive representation $\hat S = -i\p_z$ is not self-adjoint on the $z\in [0,1]$ interval and thus this approach requires extra care. The corresponding quantum spectral curve would presumably resemble more closely those in the literature as the classical limit would be an ordinary analytic spectral curve. This will be investigated elsewhere.

\clearpage

\noindent
Several other research directions  would be interesting to pursue. We list a few of them here:
\begin{itemize}
  \item One might try to directly quantize the $\mathfrak{sl}(2)$ Ruijsenaars-Schneider Hamiltonian \eqref{eq:hamipu}. This presumably involves some sort of symmetrization prescription.
  \item It would be useful to find the correct Dirac brackets for celestial variables. This would allow one to study the quantum algebra of the generators of the $SO(2,n)$ symmetry group of AdS and decide if an anomaly appears or not.
  \item It would be interesting to prove the conjecture in the Appendix about the equivalence of the closing constraints and the vanishing of the adjugate Kasteleyn matrix at two special points for strings of an arbitrary number of segments.
  \item Segmented strings can also be defined in essentially the same way on de Sitter or  BTZ backgrounds \cite{Gubser:2016wno}. It would be interesting to extend our results to these spacetimes.
  \item In this paper we studied the simplest closed string, which had only one internal degree of freedom. It would be interesting to understand how the results generalize to strings with more than four segments.
\end{itemize}
We hope to address some of these questions in the future.

\vspace{0.2in}   \centerline{\bf{Acknowledgments}} \vspace{0.2in}
I am grateful to Paolo Benincasa, Bercel Boldis, Andrea Cavagli\` a, Sergei Dubovsky, Sebasti\' an Franco,  Alba Grassi, Cristoforo Iossa, P\' eter L\' evay,  Vladimir Rosenhaus, Torben Skrzypek, Bogdan Stefanski, Alessandro Torrielli and Congkao Wen for interesting discussions and Torben Skrzypek and Alessandro Torrielli for valuable comments on the draft. I thank Jeffrey Forshaw and Ruben Sandapen for pointing out an error in an earlier version of the manuscript.
I thank the CUNY Graduate Center and New York University for hospitality.
I gratefully acknowledge discussions with the participants of the ``CFT and Holography'' meeting at Trinity College Dublin  and at the ``Integrability in Gauge and String Theory 2022'' conference at E\" otv\" os University in Budapest where some of the results were first presented.
The author has been supported by the STFC Ernest Rutherford grant ST/P004334/1.

\clearpage

\appendix

\begin{figure}[h]
\begin{center}
\includegraphics[width=6.5cm]{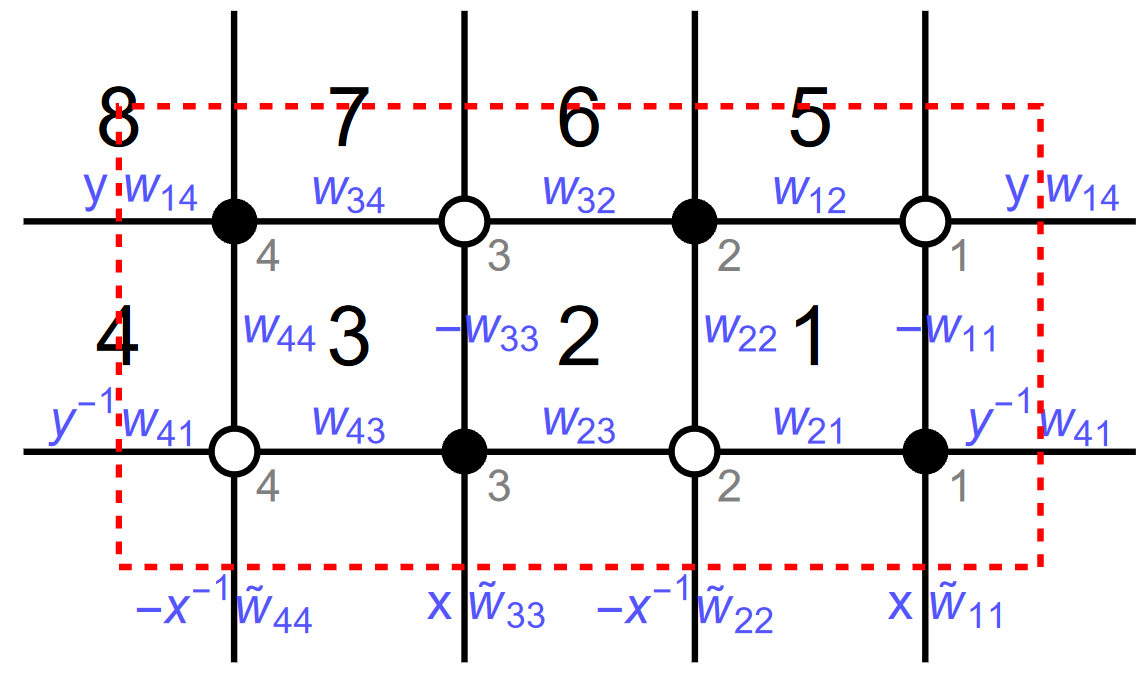}
\caption{\label{fig:y40}   Weights associated to edges in the tiling.
}
\end{center}
\end{figure}

\section{Closing constraints and the adjugate Kasteleyn matrix}

\label{sec:adj}

In this Appendix, we relate the closing constraints \eqref{eq:ccl} to properties of the so-called Kasteleyn matrix of the tiling.
The Kasteleyn matrix is the dressed adjacency matrix of the graph. Its rows and columns are labelled by the black and white nodes of the tiling and thus in the case of a four-segmented string,   it is a $4 \times 4$ matrix. Two special parameters, $x$ and $y$, multiply  those edges that cross the boundaries of the fundamental domain (red dashed lines in Figure \ref{fig:y40}). Depending on the orientation of the tiling edge (i.e. on which side of the boundary the black node lies) we multiply the weights by $x$ or $x^{-1}$  and   $y$ or $y^{-1}$   (or possibly a combination for more complicated tilings).
Finally, minus signs also have to be distributed on the edges, such that their product  around a face $F$ is given by,
\bea
\prod_{i \in \{\textrm{edges around } F\}} s_i  =
\left\{\begin{array}{l}
+1 \quad \textrm{ if } \quad  e_F=2 \mod 4 \\ %
-1 \quad \textrm{ if } \quad  e_F=0 \mod 4
 \end{array}\right. \, ,
\eea
where $e_F$ denotes the number of edges around $F$. Such an assignment of minus signs is always possible.
For a detailed discussion, we refer the reader to \cite{Vegh:2021jqo}.

Let us now focus on the square tiling that is relevant for the four-segmented string and  assign 16 unfixed weights to its edges as in Figure \ref{fig:y40}. The Kasteleyn matrix is easy to compute,
\be
  \nonumber
K = \left( \begin{array}{cccc}
x \tilde w_{11}-w_{11} & w_{12} & 0 &  y w_{14} \\
w_{21} & w_{22}-x^{-1}\tilde w_{22} & w_{23} & 0 \\
0 & w_{32} & x \tilde w_{33}-w_{33} & w_{34} \\
y^{-1}w_{41} & 0 & w_{43} & w_{44}-x^{-1}\tilde w_{44}
\end{array} \right) \, .
\ee
For each function on the nodes, one can associate a gauge transformation on the edge weights in the following way. At a white (black) node the transformation multiplies (divides)
the weights of all edges incident to that node by the function value. Edge weights related
by gauge transformations are equivalent. Note that constant functions give trivial gauge transformations, because each edge weight is divided and multiplied by the same number. In the four-segmented case, there are 16 tiling edges, 8 nodes and therefore a 7d space of gauge transformations. This leaves us with $16-7=9$ independent variables before demanding that the closing constraints are satisfied.

Note that in a certain gauge the edge weights are simply given by the $x_i, y_i$ variables, which in this gauge appear on the edges according to Figure~\ref{fig:lego0} (left). The spectral curve can also be computed using the Kasteleyn matrix by taking the determinant,
\be
  \nonumber
   \det K(x,y) = 0 \, ,
\ee
which gives the same result as before in \eqref{eq:spec4b} and \eqref{eq:spec4bb}.

Although we already have the expression \eqref{eq:ccl} for the closing constraints, we will now make an observation, which transforms the constraints into a form that is more natural in the language of the tiling and could perhaps generalize to other cases.
For this we will need the definition of the adjugate of the Kasteleyn matrix,
\be
  \nonumber
  \textrm{adj}\ K := K^{-1} \, \det K  \, .
\ee
The observation is that the  constraints are equivalent to demanding that the adjugate matrix vanishes at two special points
\setlength\fboxsep{0.3cm}
\be
  \label{eq:adjeq}
 \boxed{ \quad  \textrm{adj}\ K = 0  \, , \qquad \textrm{at} \ x=\pm 1, \ y=-1 \, . \ \ }
\ee
A lenghty, but straightforward calculation shows that demanding \eqref{eq:adjeq} gives precisely seven constraints on the edge weights, which finally gives a $9-7=2$ dimensional phase space for the four-segmented string. One of the constraints fixes the normalization of the spectral parameter, while the other six are equivalent to the constraints in \eqref{eq:ccl}. We have checked this explicitly for the 4-segmented string and (numerically) for the 6-segmented string. It is tempting to conjecture that the statement is also true for strings with an arbitrary number of segments.

\clearpage

\bibliographystyle{JHEP}
\bibliography{paper}

\end{document}